\documentclass[a4paper,11pt]{article}

\pdfoutput=1
\usepackage{jheppub}

\usepackage{graphicx,mathtools,caption,subcaption,footmisc}

\def\zh{\zeta}
\def\DS{\Delta S_E}
\def\euv{\epsilon_{{\text{uv}}}}

\title{Holographic entanglement entropy for relativistic hydrodynamic flows} 

\author[a]{Jyotirmoy Bhattacharya }
\affiliation[a]{Department of Physics, Indian Institute of Technology Kharagpur, Kharagpur 721302, India.}
\author[b]{, Parthajit Biswas }
\affiliation[b]{Department of Physics, Ramakrishna Mission Vivekananda Educational and Research Institute, Belur Math, Howrah 711202, India.}
\author[c]{, A. Chandranathan }
\affiliation[c]{International Centre for Theoretical Sciences (ICTS-TIFR), Tata Institute of Fundamental Research, Shivakote, Hesaraghatta, Bangalore 560089, India.}
\author[a]{, and \\ Sayan Kumar Das }

\emailAdd{chandranathan.a@icts.res.in, jyoti@phy.iitkgp.ac.in, parthajitbiswas8@gmail.com,
sayankumardas@iitkgp.ac.in.}

\abstract{We study the behaviour of holographic entanglement entropy (HEE) in near equilibrium thermal states which are macroscopically described by conformal relativistic hydrodynamic flows dual to dynamical black brane geometries. We compute HEE for strip-shaped subsystems in boundary dimensions $d=2,3,4$, which provides us with general qualitative inferences on the interplay between fluid flows and entanglement dynamics. At first, we consider the zeroth order in hydrodynamic derivative expansion, holographically described by stationary boosted black branes. Working non-perturbatively in fluid velocity, we find that, as the fluid velocity approaches its relativistic upper limit, the UV regulated HEE exhibits a divergence at arbitrary temperature. Also, the holographic mutual information between two relatively close subsystems vanishes at some critical fluid velocity and remains zero beyond it. We then compute HEE in an excited state of the fluid in the presence of the sound mode. As a simplified setup, we first work with non-dissipative dynamics in $d=2$, where the time evolution of HEE is studied in the presence of the sound mode and a propagating pressure pulse. In $d= 4$, working upto first order in derivative expansion, we find that dissipative sound modes produce an additional dynamical UV divergence which is subleading compared to the `area law divergence'. No such divergence is observed for dissipative sound mode in $d=3$.}

\arxivnumber{2211.14271}

\begin{document}

%
%
\maketitle
\flushbottom
%

\section{Introduction}\label{sec:intro}

%
The translationally invariant macroscopic thermal dynamics of a quantum field theoretic system is believed to be 
captured by a set of effective hydrodynamic degrees of freedom. The theory governing these hydrodynamic
variables is primarily based on the underlying symmetries, which makes its applicability significantly universal. 
In the hydrodynamic description, the relevant microscopic information is packaged into a few parameters 
- the transport coefficients. The fluid variables  capture
the near equilibrium dynamics very efficiently, while the corresponding microscopic dynamics is extremely complex. 
The intricate interplay between the macro and micro dynamics usually provides us with enourmous insights 
into a physical system, but usually is a difficult question to address. 
The difficulty mainly arises due to the intractability of the microscopic degrees of freedom in such a situation. 
The fluid-gravity correspondence \cite{Bhattacharyya:2007vjd} provides us with a very 
special setting where a complete picture is provided and the transport coefficients are determined from the dynamics of a dual black brane. In this picture, the various length scales of the hydrodynamic system are encoded in a holographic direction. 

The fluid-gravity correspondence maps the dynamics of a $d$ dimensional conformal relativistic fluid to that of 
a AdS$_{(d+1)}$ black brane. The microscopic description of this holographic fluid is provided by a strongly coupled 
conformal gauge theory. The black brane has a complete knowledge of this strongly coupled microscopics. 
This is why, all the hydrodynamic transport coefficients can be computed from the black brane geometry, in a regime 
where a direct field theoretic calculation is intractable. Similarly, the holographic geometry can be used to extract 
other microscopic information about the near-equilibrium thermal states. Particularly, 
the entanglement structure of these underlying quantum field theoretic states, which are evolving towards equilibrium, is significantly interesting. An excellent measure of this entanglement is provided by the holographic entanglement 
entropy (HEE) \cite{Ryu:2006bv, Ryu:2006ef, Hubeny:2007xt, Nishioka:2009un, Rangamani:2016dms, Headrick:2019eth}. In this paper, 
our objective is to apply the holographic dictionary to study the dynamics of HEE for the states 
whose macrodynamics corresponds to specific hydrodynamic flows. 

\begin{figure}[ht]
\centering
\includegraphics[width = 0.9\textwidth]{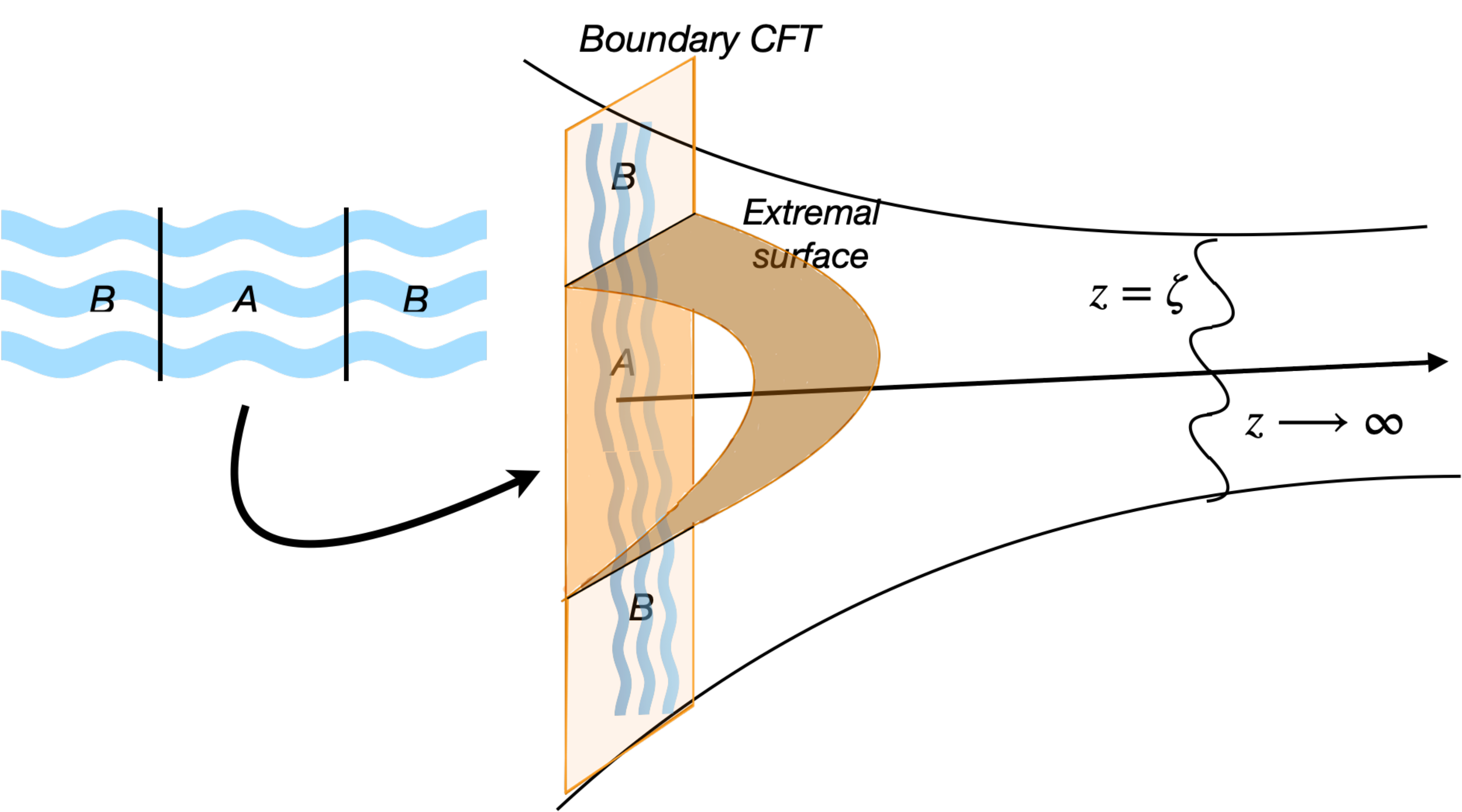}
\caption{A Schematic representation of the strip-shaped subsystem $A$ on a canonical time-slice of the AdS boundary for which holographic entanglement is to be computed. 
The underlying state is a fluctuating fluid configuration in our set-up.}\label{fig:Scheme}
\end{figure}

In order to compute entanglement entropy in a continuum field theoretic set up, we first choose 
a subsystem $A$, usually taken to be a space-like sub-region on a chosen time-slice. 
The entanglement entropy of a given state, is then given by the Von Neumann entropy of the reduced density matrix 
obtained by integrating out the degrees of freedom outside $A$. The knowledge of this entanglement entropy 
for various subsystems provides us with enourmous quantum information about the underlying state. Holographically, 
this entanglement entropy is computed by the celebrated Ryu-Takayanagi formula \cite{Ryu:2006bv, Ryu:2006ef}, which was generalized to a more general covariant form in \cite{Hubeny:2007xt}. 
In this prescription, the HEE for $A$ is given by 
\begin{equation}\label{EntFormula}
 S_A = \frac{ A_\Gamma}{4 G_N},  
\end{equation}
where $A_\Gamma$ is the area of the co-dimension 2 surface 
homologous to $A$ which has been extremized on the bulk geometry dual to the 
the field theoretic boundary state. The subscript $\Gamma$ refers to this extremal surface.
In our work here, using the fluid-gravity map, 
we will explicitly write down the geometries dual to specific hydrodynamic configurations. Subsequently, 
we shall compute the area of the extremal surface in these geometries to obtain HEE ($S_A$) for the hydrodynamic states. 
For the subsystem $A$, we choose a strip-shaped sub-region on a canonical time slice of the AdS boundary. Our choice 
of subsystem is time-independent, while the underlying state is a fluctuating fluid configuration 
(see fig.\ref{fig:Scheme}).
%
%
%

%
In this paper, we shall primarily focus on two varieties of fluid configurations. In the first 
case, we shall consider steady state flows where the fluid is moving with a constant velocity (see section  \ref{sec:bbb}). 
The computation of HEE for such thermal systems has been attempted by several authors in the recent past
\cite{Fischler:2012uv, Fischler:2012ca, Bhattacharya:2013bna, Blanco:2013joa, Mishra:2015cpa, Mishra:2016yor, Bhatta:2019eog, Maulik:2020tzm, ChowdhuryRoy:2022dgo}. 
However, in all these earlier works, some simplifying assumption have been made, and therefore, a thorough analysis,  
particularly in the regime where the fluid velocity approaches its relativistic upper bound, seems to be missing so far. 
For this steady state case, apart from HEE, we also compute the holographic mutual information between 
two non-overlapping strip-shaped subsystems. Among various other 
results, we find that there is a critical fluid velocity beyond which mutual information vanishes due to an exchange of 
dominance of extremal surfaces. This is similar to the `disentangling transition' previously observed in 
\cite{Headrick:2010zt,Fischler:2012uv}, where the abrupt vanishing of mutual information was observed as the separation between the two subsystems in question exceeded a certain critical value
\footnote{This abrupt vanishing of mutual information at a finite separation between the subsystems, is understood to be a feature of large central charge and has been observed in all previous similar holographic studies (see \cite{Headrick:2010zt,Headrick:2019eth} for a more detailed discussion on this point).}. In this paper, we observe such a `disentangling transition' taking place 
as a function of the fluid velocity. This we believe, is a new and important result, which was not possible to obtain 
without the non-perturbative approach of our paper.

In another case, we consider fluid configurations where a propagating sound wave
have been set up with some initial conditions. We then proceed to study the change 
in HEE produced due to such fluid fluctuations in the linearized approximation, assuming a small amplitude of the excitation 
(see section \ref{sec:hydromodes}). Since, sound waves constitute
one of the basic fluid excitations, it is interesting to understand the HEE dynamics associated with it. 
In $d=2$ the sound mode is non-dissipative, while in higher dimensions these fluctuations 
dissipate away to equilibrium. In the latter case, we aim to capture the time-evolution of HEE as the fluid settles 
back to equilibrium. In $d=2$ where the sound mode is non-dissipative, we can superpose them to 
constitute a pressure or energy density pulse. In this case, we study how HEE evolves with time 
as the pulse passes across the subsystem. 
Our investigation here is reminiscent of \cite{Narayan:2012ks}
where HEE has been computed in the AdS plane wave geometry with an energy flux flowing through 
the subsystem. We are also inspired by the body of work on the study of HEE following quantum quenches (see for instance
\cite{Nozaki:2013wia, Rangamani:2015agy, Jahn:2017xsg}).
In fact, we may consider our fluid dynamical set-up to be a thermal state which has been quenched by a macroscopic observable 
associated with the energy-momentum tensor, and we study the subsequent late time evolution of HEE.

In our work here, we employ a mixture of analytical and numerical techniques. In particular, most of our 
results in $d=2$ is analytical. We use this analytical result to optimize our numerical procedure, which are 
then applied to $d=3,4$. The qualitative inferences which we draw from our results in 
$d=2,3,4$ may be straightforwardly generalized to higher dimensions. 

\subsection*{Summary of results}
Before proceeding to the details, let us summarize the important new results of our paper below.
In this paper we investigate the behaviour of entanglement entropy in a class of conformal fluid states, using holographic techniques, for a strip like subsystem. 
\begin{itemize}
\item We compute HEE for boosted black holes in $d+1$ ($d=2,3,4$), viewed as holographic geometries dual to fluids moving with 
a constant velocity. Our subsystem lies on a `canonical time slice' and its size is fixed as we vary the fluid parameters.  
Our calculation is performed without making any assumption on temperature or velocity parameter. All earlier studies of this system had made some simplifying assumption about these parameters and lifting those assumptions is a new aspect of our study. In particular, we curiously observe that the regulated HEE diverges as the fluid velocity approaches its relativisitc upper bound ($v \rightarrow 1$) (see fig.\ref{fig:2dnumamatch} and fig.\ref{fig:highDboost}). See section \ref{sec:bbb} for more details.  
\item For the constant velocity fluid flows, we also compute holographic mutual information. We find that this quantity undergoes an abrupt `disentangling' transition for some critical velocity for any separation between the subsystems (see fig.\ref{fig:MI2d} and fig.\ref{fig:MIhighD}). This critical velocity tends to 1 as the separation between the subsystem tends to zero. The existence of this critical velocity is a new and interesting result. 
See section \ref{sssec:HMIbb2D} and section \ref{ssec:bbinD} for more details. 
\item For holographic fluids in d=2, we have studied the time dependence of HEE as a non-dissipative propagating pressure pulse passes 
through the demarcated subsystem. The geometric dual to this pressure (or energy) pulse is an exact solution to the bulk Einstein equations. We find that HEE has a curious dip when the pulse enters the subsystem, followed by an expected rise and plateau while the pulse resides inside the subsystem (see fig.\ref{fig:delSpvst}). We report these results in section \ref{sound2D}. 
\item For holographic fluids in $d=3,4$, we study the behaviour of HEE in the presence of a dissipative sound mode. In this case HEE has the usual expected decay closely following the damping of the sound mode. However, surprisingly, we find that in d=4, there is an additional UV divergence which is subdominant compared to usual area law (see \eqref{delSform4d}). We present the details of these results in section \ref{ssec:sound4D}. 
\end{itemize}

\section{Entanglement in steady state relativistic fluid flows}\label{sec:bbb}
%
\subsection{General Procedure}\label{ssec:bbbgen}
%
%
Before proceeding with the hydrodynamic excitations, we would like to consider the simple steady state case, when 
the fluid is moving with a constant velocity. More precisely, we consider 
a relativistic conformal fluid, with a constant velocity, in a flat Minkowski background with $$\eta_{\mu \nu} = \text{Diag}\{-1,1,1, \dots\}.$$ Holographically, this fluid configuration is described by a bulk metric of a boosted black brane, which  
in $d+1$ dimensional Schwarzschild coordinates
\footnote{In this section, we use the Schwarzschild coordinates since this has been predominantly 
used in the literature on HEE. The derivative expansion 
in the fluid-gravity correspondence \cite{Bhattacharyya:2007vjd} necessitate the use of more regular Eddington-Finklestein (EF) coordinates, so as to formulate a well defined derivative expansion. 
Therefore, while discussing hydrodynamics modes with first order derivative corrections (in section  \ref{sec:hydromodes}) we will use EF coordinates and  in appendix \ref{app:EFcoord} we have explicated  
the transformation between the two coordinates.}, takes the form
\footnote{We shall set the AdS radius $\mathcal R_{AdS} = 1$ throughout this paper. \label{foot:Rads}}
\begin{equation}\label{schboost}
 ds^2_{\text{Sch}} = \frac{1}{z^2} \left( -  dt^2 + dx^2 + g(z) \big( dt ~- ~v ~dx \big)^2 
 + \frac{dz^2}{f(z)} + \sum_{i = 1}^{d-2} dy_i^2 \right) 
\end{equation}
where 
\begin{equation*}
 f(z) = 1 - \left(\frac{z}{\zh}\right)^d,~~ g(z) =  \gamma^2 \left(\frac{z}{\zh}\right)^d ,~~ \gamma = \frac{1}{\sqrt{1-v^2}}. 
\end{equation*}
Here, $0 \leq v \leq 1$ is the boost parameter which is identified with the fluid velocity. Also, the AdS boundary is reached in the limit 
$z \rightarrow 0$. We denote the horizon radius with  
$\zh$, which is related to the temperature of the black brane as 
\begin{equation}\label{temp}
  T = \frac{d}{4 \pi \zh} \equiv \frac{1}{\beta}. 
\end{equation}
This is also identified with the steady state temperature of the fluid. 

The equations of hydrodynamics in steady state is provided by 
the conservation of the energy-momentum tensor, which has the following constitutive relations  
\begin{equation}\label{steadyperflu}
 T^{\mu \nu} = \mathcal E(T) u^\mu u^\nu + \mathcal P (T) P^{\mu \nu} + \mathcal O (\partial ),
\end{equation}
where $P^{\mu \nu} = \eta^{\mu \nu} + u^\mu u^\nu$ is the projector orthogonal to 
the fluid velocity $u^\mu$, $\mathcal E(T)$ and $\mathcal P(T)$ are 
respectively the energy and pressure densities, while $\mathcal O (\partial )$ represents 
higher derivative corrections to this perfect fluid form. 
In steady state, the temperature and fluid velocity are constants over 
space-time, implying that all the higher derivative terms vanish. Also, 
for such a steady state it is immediately clear that 
the fluid variables satisfy the conservation equations $\partial_\mu T^{\mu \nu}=0$. 
The form of the energy momentum tensor \eqref{steadyperflu} follows 
from the boosted AdS black brane \eqref{schboost}, via the AdS-CFT dictionary. 
The duality also determines the precise functional forms
\begin{equation}\label{PEperflu}
 \mathcal P (T) = \frac{1}{16 \pi G_N} \left( \frac{4 \pi T}{d}\right)^d , ~~ \mathcal E (T) = (d-1)\mathcal P (T)
\end{equation}

On the boundary of the boosted black brane, we choose a strip like 
subsystem ($A$) on a canonical time-slice\footnote{Our boundary time-coordinate is chosen to be the boundary limit 
of bulk Eddington-Finkelstein-time or Schwarzschild-time, both being 
identical on the boundary of AdS. Constant time-slices on the boundary, with the time coordinate defined in this way, will be referred to as `canonical time-slice'. Note that, since the Eddington-Finkelstein coordinates are indispensable for the fluid-gravity bulk metric \cite{Bhattacharyya:2007vjd}, the boundary time coordinate defined in this way, is somewhat special for the holographic hydrodynamics. \label{cantmdef}} and we wish to quantify the entanglement of $A$ with the region outside the strip. The underlying fluid is moving with a constant velocity in some given direction (see fig.\ref{fig:Scheme} for a schematic of the situation).

For zero boost ($v=0$), the HEE for black branes in arbitrary dimensions have been 
computed in \cite{Fischler:2012uv, Fischler:2012ca, Bhattacharya:2013bna}. 
The extension of this analysis to boosted black branes have been 
attempted in several papers earlier \cite{Blanco:2013joa,Mishra:2015cpa,Mishra:2016yor,Bhatta:2019eog,Maulik:2020tzm,ChowdhuryRoy:2022dgo}. However, in all these papers some simplifying approximations have been employed, such as small boost in \cite{Mishra:2015cpa, Mishra:2016yor, Bhatta:2019eog}, or small temperature \cite{Blanco:2013joa}. 
To our knowledge, a complete computation of HEE for arbitrary boost and arbitrary temperature seems to be missing in the literature. This subsection aims to perform a thorough analysis of this situation, particularly focusing on the situation when the fluid velocity approaches its relativistic upper bound ($v \rightarrow 1$) at an arbitrary  background temperature.

As pointed out in \cite{Blanco:2013joa}, for a non-perturbative HEE computation in this scenario, 
the extremal surface is not expected to lie on a bulk time-slice. Therefore, even if we are 
using Schwarzschild coordinates \eqref{schboost}, for the boosted case, we must employ the covariant 
protocol of \cite{Hubeny:2007xt}.

Here, we shall consider a strip-shaped subsystem in a $d$ dimensional space-time. The symmetries of such a strip-like subsystem simplify some of the calculations significantly.  From the boundary point of view, there are two special spatial directions in this set-up.  The short edge of the strip is directed along some spatial direction, while the boosted fluid may be flowing in any other independent direction. The result of HEE is expected to depend on the angle between these two vectors. But again for the sake of simplicity, in this paper, we will assume these directions to be collinear, i.e. the short edge of the strip is extended along the same direction as the boost. With this in mind, the embedding ansatz for the extremal surface, which is a co-dimension 2 surface homologous to our strip like subsystem on the boundary (see fig.\ref{fig:Scheme}), is taken to be 
\begin{equation}\label{embedgend}
 \Gamma: ~x= x(\phi, \tilde y_i),~ t=t(\phi, \tilde y_i),~ z = z_\star \cos \phi, ~ y_i = \tilde y_i. 
\end{equation}
In other words, the world-volume of our co-dimension 2 extremal surface is parameterized 
by $\phi$ and $\tilde y_i$. The $\phi$-coordinate on the world-volume takes values between 
$-\pi/2 \leq \phi \leq \pi/2$, and $z_\star (< \zh)$ is the turning point of the extremal surface, which will be determined in terms of the width $\ell$ of the strip on the boundary. 
Further, on the boundary, to regulate the infinite volume arising due to the long sides of the strip, we will
consider the $\tilde y_i$ coordinates to be bounded  
$$ -\frac{L}{2} \leq \tilde y_i \leq \frac{L}{2}, ~~ i = 1,2 \cdots d-2, ~~ \text{with} ~~L \rightarrow \infty.$$
Now, since the boost is along the $x$-direction (along the short-edge of the strip) \eqref{schboost}, we retain 
translation invariance along the $\tilde y_i$-directions. Thus, 
we need to solve for two functions of one variable, $x(\phi)$ and $t(\phi)$. 
The area functional of $\Gamma$, to be extremized, is given by 
\begin{equation}\label{schbooAform}
A_\Gamma = L^{d-2} \int d \phi \left( \frac{1}{z_\star \cos \phi }\right)^{d-1} \left[ -(1-g) t'^2 + (1+ v^2 g) x'^2 - 2 v g t' x' + \frac{z_\star^2 \sin^2 \phi}{f} \right]^{\frac{1}{2}}. 
\end{equation}
where $x'$ and $t'$ denotes derivatives of these functions with respect to $\phi$. 
Here, the factor of $L^{d-2}$ arises due to the integrals 
over the $\tilde y_i$ coordinates on the world-volume of the extremal surface. Also, in \eqref{schbooAform} $g = g(z_\star \cos \phi)$, and $f = f(z_\star \cos \phi)$  ~  and their functional forms are given in \eqref{schboost}. 
The equation of motion for $x(\phi)$ and $t(\phi)$ which follows from the extremization of 
$A_{\Gamma}$ is given by
\begin{equation}\label{scheom}
\begin{split}
& \frac{\partial}{\partial \phi} \left( \frac{(1-g) t' + v g x'}{\mathcal K \left(\phi, x(\phi), t(\phi)\right)} \right) = 0,  ~~~
 \frac{\partial}{\partial \phi} \left( \frac{(1 + v^2 g) x' - v g t'}{\mathcal K \left(\phi, x(\phi), t(\phi)\right)} \right) = 0, 
 ~~\text{where,}
\\
&\mathcal K \left(\phi, x(\phi), t(\phi)\right) = (z_\star \cos \phi)^{d-1} \left[ -(1-g) t'^2 + (1+ v^2 g) x'^2 - 2 v g t' x' + \frac{z_\star^2 \sin^2 \phi}{f} \right]^{\frac{1}{2}}.
\end{split}
\end{equation}
We now have to solve these equations \eqref{scheom}, with the following boundary conditions 
\begin{equation}\label{scheombdy}
 x(-\pi/2) = - \frac{\ell}{2}, ~ x(\pi/2) = \frac{\ell}{2},~~t(-\pi/2) = 0 = t(\pi/2).
\end{equation}
Since, we are dealing with a stationary scenario, we expect the HEE to be independent of time. Hence 
without any loss of generality we have chosen the boundary time slice at $t_b=0$. 

The equations \eqref{scheom}, \eqref{scheombdy} is difficult to solve analytically. Therefore, we have to resort to numerical analysis for computing the area of the extremal surface. 

In the absence of boost ($v=0$), $t(\phi) = \text{constant}$ is a solution to the equations \eqref{scheom}. This implies 
that, in the unboosted case, if we use Schwarzschild coordinates, the extremal surface always lies on a constant time-slice 
throughout the bulk. This, however, is not the case in the presence of boost, even if we work in Schwarzschild coordinates. 
%
%

\subsubsection*{Extremal surface reparameterization: $\phi \rightarrow z$}
It is often practically useful to consider the embedding of the extremal surface such that 
the AdS-radial coordiate $z$ is identified with one of the world-volume coordinates. With such 
a choice, the embedding equations for the strip-subsystem would read 
\begin{equation}\label{embedgendz}
 \Gamma: ~x= x(z),~ t=t(z),~ z = z, ~ y_i = \tilde y_i, 
\end{equation}
with the equations of motion \eqref{scheom} being minimally modified to 
\begin{equation}\label{scheomz}
\begin{split}
& \frac{\partial}{\partial z} \left( \frac{(1-g) t' + v g x'}{\mathcal K \left(z, x(z), t(z)\right)} \right) = 0,  ~~~
 \frac{\partial}{\partial z} \left( \frac{(1 + v^2 g) x' - v g t'}{\mathcal K \left(z, x(z), t(z)\right)} \right) = 0, 
 ~~\text{where,}
\\
&\mathcal K \left(z, x(z), t(z)\right) = z^{d-1} \left[ -(1-g) t'^2 + (1+ v^2 g) x'^2 - 2 v g t' x' + \frac{1}{f} \right]^{\frac{1}{2}}.
\end{split}
\end{equation}
Now $x'$ and $t'$ denote derivatives with respect to $z$, while $f=f(z)$ and $g=g(z)$ as given in \eqref{schboost}. In this case, the 
boundary conditions \eqref{scheombdy}, are implemented with a minor difference. The solution to \eqref{scheomz} has two distinct branches
($x_{\pm}(z), t_{\pm}(z)$) which are valid in the domain $z \in [z_\star, 0]$, with $z_\star$ being the turning point. 
The boundary conditions for the two roots are implemented in the following way 
\begin{equation}\label{scheombdyz}
 x_{\pm} (z \rightarrow 0) = \pm \frac{\ell}{2}, ~~t_{\pm}(z \rightarrow 0) = 0, ~~x'_{\pm}(z \rightarrow z_\star) \rightarrow \infty, 
 ~~ t'_{\pm}(z \rightarrow z_\star) \rightarrow \infty.
\end{equation}
The two branches are smoothly stiched together at the turning point $z_\star$ (see fig.\ref{fig:extreme2D} for the extremal line in $d=2$). 

The area functional, after this reparameterization, is given by 
\begin{equation}\label{schbooAformz}
A_\Gamma = L^{d-2} \int \frac{dz}{z^{(d-1)}} \left[ -(1-g) t'^2 + (1+ v^2 g) x'^2 + 2 v g t' x' + \frac{1}{f} \right]^{\frac{1}{2}}. 
\end{equation}
This area functional evaluated on the extremal surface gives the HEE by the formula \eqref{EntFormula}. We shall regulate the UV divergence  by removing the HEE corresponding to pure AdS 
\begin{equation}\label{Sedefb}
 \mathcal S = \frac{A_\Gamma}{4 G_N}, ~~\Delta \mathcal S = \mathcal S - S^{\text{vac}} \eqqcolon \frac{L^{(d-2)}}{4 G_N} \Delta S_E.
\end{equation}
The  $\Delta S_E$, as defined above, will be useful for quoting our numerical results.

\subsection{Fluid flows in $d=2$ dual to boosted BTZ black brane}\label{ssec:bBTZ}

In $1+1$D boundary, an exact expression of HEE for the steady state fluid flow dual to boosted BTZ black brane, may be obtained 
by a simple argument using a result available in \cite{Kusuki_2017}. 
After presenting this argument in section \ref{sssec:HEEbb2D}, we proceed directly to
solve the equations \eqref{scheom} numerically and verify our argument. 
Further, in $d=2$ we can also use our analytical result to compute 
mutual information for configurations with constant fluid velocity, which we present 
in section  \ref{sssec:HMIbb2D}.

Our numerical procedure in $d=2$, having been verified by an exact analytical result, 
is then straightforwardly carried over to higher dimensions in section  \ref{ssec:bbinD}. 
%

\subsubsection{Holographic Entanglement entropy}\label{sssec:HEEbb2D}
%
\subsubsection*{Analytical exact result}
%
%
%
Let us first consider the unboosted BTZ black brane in Schwarzschild coordinates
\begin{equation} \label{btzubsch}
ds^2 = \frac{1}{z^2} \left( - f(z) dt^2 + \frac{1}{f(z)} dz^2 + dx^2 \right). 
\end{equation} 
where, $f(z)$ is given by the same function in \eqref{schboost} for $d=2$. This metric is dual 
to a static fluid configuration in $d=2$ with fluid velocity $u^\mu = \{ -1,0\}$ and temperature given by \eqref{temp}. 

Now, we wish to define our boundary subsystem ($A$) as the region between two space like points, say $P$ and $Q$. Without the loss of generality, we can consider the boundary points to be $P \equiv (0,0)$ and $Q \equiv (\tau, \xi)$. The HEE for this situation was worked out in \cite{Hubeny:2007xt, Kusuki_2017} 
and the answer is given by 
\begin{equation} \label{difslres}
\mathcal S_A = \frac{c}{6} \ln \left( \frac{\beta^2}{\pi^2 \euv^2} \sinh \left( \frac{\pi}{\beta} \left( \xi + \tau \right)  \right) \sinh \left(  \frac{\pi}{\beta} \left( \xi - \tau \right)  \right)   \right) 
\end{equation} 
where $\euv$ is the ultraviolet cut-off 
\footnote{Also see \cite{Cadoni:2010ztg, Hartman:2013qma, Caputa:2013eka} for other related results in the context of the BTZ black hole.}. 

Let us boost this set-up with velocity $v$, so that the fluid velocity becomes $u^\mu = \gamma (-1, v)$. Under this boost, 
the bulk BTZ metric \eqref{btzubsch} goes over to \eqref{schboost} (for $d=2$).  
Besides affecting the underlying fluid state, the boost also transforms the end points of our subsystem $P$ and $Q$. 
Consequently, our subsystem $A$ will now lie on a different time-slice compared to the time-slice before boosting. 
We wish to compute HEE for the boosted BTZ when our subsystem $A$ lies on a canonical time-slice (see footnote \ref{cantmdef}). 
\begin{figure}[htb]
\centering
\includegraphics[width = 0.8\textwidth]{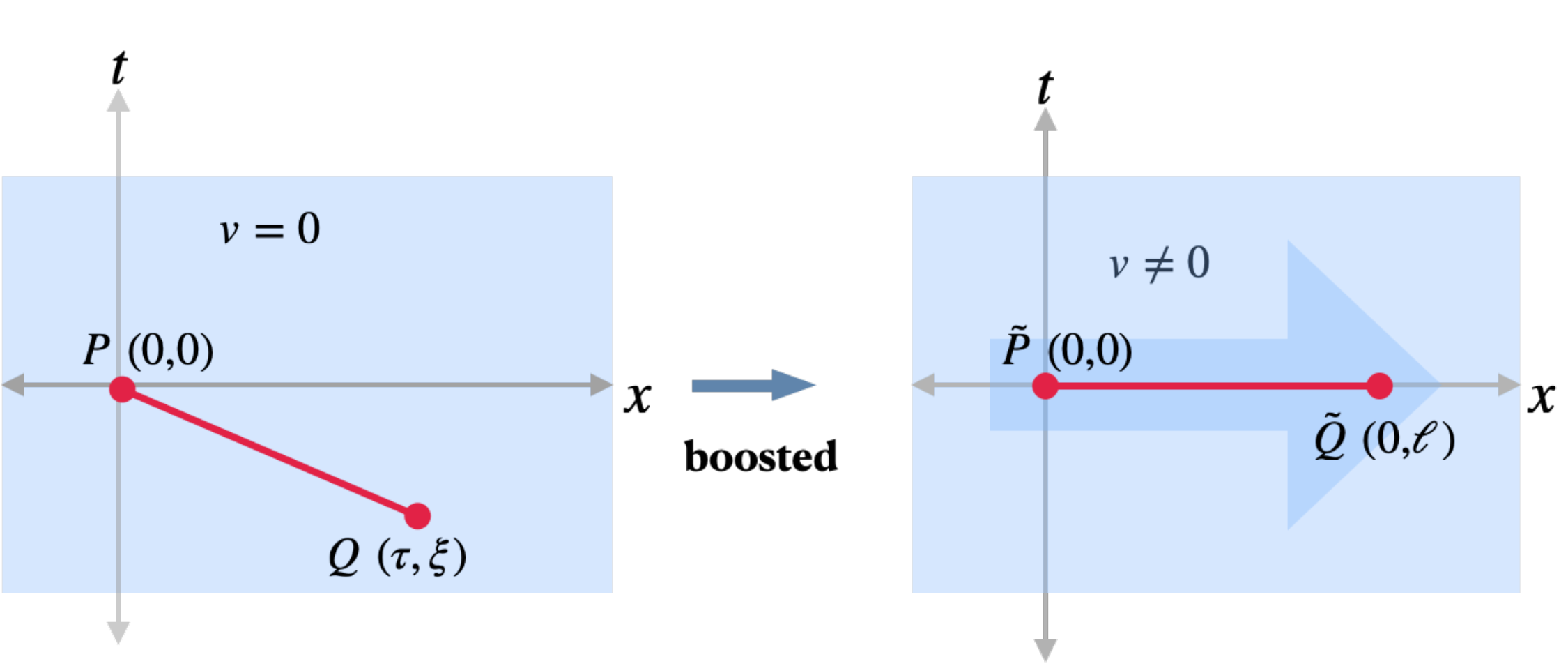}
\caption{Schematic representation of the argument leading to the computation of HEE for the steady state of a $1+1$D conformal fluid, 
moving with a constant velocity. We start with the static fluid and find HEE for a subsystem defined by the points $P$ and $Q$ as shown in the left figure. This situation is boosted so that  the system, now defined by the points $\tilde P$ and $\tilde Q$, returns back on to a canonical time slice as shown in the right figure (see footnote \ref{cantmdef}). The boosting ensures that the background fluid is moving with a constant velocity, equal to the value of the boost parameter. This procedure enables us to write down an exact expression for HEE in this case, as given by \eqref{boostres}.}\label{fig:boostflow}
\end{figure}

The boosted coordinates would be related to the old coordinates by the transformation
\begin{equation} 
\tilde t = u^\mu x_\mu = \gamma \left( t + v x \right) , ~~\tilde x = \sqrt{x^\nu x^{\mu} P_{\nu \mu} } = \gamma \left( x + v t \right)  \, 
\end{equation} 
where $P_{\mu \nu} = \eta_{\mu \nu} + u_\mu u_\nu$ is the projector orthogonal to $u^\mu$. Now, 
prior to boosting if the end points of $A$ are $P$ and $Q$, then after boosting they must transform to $\tilde P = (0,0)$ and $\tilde Q = (0, \ell)$, which will ensure that $\tilde P$ and $\tilde Q$ both lie on the canonical time-slice. Thus, we get back to a canonical time-slice only after boosting. This would ensure that the underlying state is that of a fluid moving with a constant velocity, while we are measuring entanglement on a canonical time slice (see fig.\ref{fig:boostflow}). To accomplish this we choose $(\tau, \xi)$ so that 
\begin{equation} 
 \tilde t^{\tilde Q} = \gamma \left( \tau + v ~\xi \right) = 0,  ~~ \tilde x^{\tilde Q} = \gamma \left( \xi + v ~\tau \right) = \ell, 
\end{equation}
where $ \{ \tilde t^{\tilde Q} , \tilde x^{\tilde Q}\}$ are the coordinates of the point $\tilde Q$ which is the boost transformed 
version of point $Q = \{ \tau, \xi\}$. This fixes $\tau$ and $\xi$ to following 
\begin{equation} \label{xitaus}
\xi + \tau = \alpha \ell,~ \xi - \tau = \frac{1}{\alpha} \ell , ~~\text{where} ~\alpha = \left( \frac{1 - v}{1+ v} \right)^{\frac{1}{2}} .
\end{equation}
Plugging this back into \eqref{difslres}, we obtain 
\begin{equation} \label{boostres}
\mathcal S_A (\ell , v) = \frac{c}{6} \ln \left( \frac{\beta^2}{\pi^2 \euv^2} \sinh \left( \frac{\pi \alpha \ell}{\beta}  \right) \sinh \left(  \frac{\pi \ell}{\alpha \beta}  \right)   \right) 
\end{equation} 
This is the final expression of HEE for the $1+1$D hydrodynamics state, where the fluid is moving with a constant velocity. 
In this next subsection, we shall reproduce this numerically directly from the equations \eqref{scheom}, \eqref{scheombdy}. 

Note that the result for this boosted case \eqref{boostres} is significantly different from the unboosted thermal state for the same subsystem lying on a canonical time slice. The later is given by setting $\alpha = 1$ in \eqref{boostres}
\begin{equation} \label{boostresnob}
\mathcal S_A (\ell, v =0) = \frac{c}{3} \ln \left( \frac{\beta}{\pi \euv} \sinh \left( \frac{\pi \ell}{\beta}  \right)   \right) 
\end{equation} 

Although trivial, it is also interesting to apply the logic of this subsection for the vacuum state. 
The boundary vacuum state is dual to pure AdS, for which the analogue of the formula \eqref{difslres} applied to an identical subsystem $A$ with boundary points $P$ and $Q$, is given by 
\begin{equation} \label{vacent}
\mathcal S_A^{(\text{vac})} = \frac{c}{6} \ln \left( \frac{\xi^2 - \tau^2}{\euv^2}  \right) 
\end{equation} 
If we plug \eqref{xitaus} back into \eqref{vacent} we get 
\begin{equation} \label{vacent2}
\mathcal S_A^{(\text{vac})} = \frac{c}{3} \ln \left( \frac{\ell}{\euv}  \right) 
\end{equation} 
which is identical to HEE for unboosted vacuum. We find this match simply because the vacuum is 
boost invariant. On the other hand, the boundary state that corresponds to a black brane is dual to static fluid (deconfined plasma). Such a state is not boost invariant. Viewed as hydrodynamic states, the fluid at rest is different compared to the fluid moving with a constant velocity and their structure of entanglement is also distinct. 

\subsubsection*{Large velocity limit ($v \rightarrow 1, \gamma \rightarrow \infty$)}
It is instructive to explicate the large velocity limit of the exact analytical formula \eqref{boostres}. If we first perform 
a small temperature (large $\beta$) expansion of \eqref{boostres} we get 
\begin{equation}\label{bosmlT}
 \mathcal S_A =  \mathcal S_A^{(\text{vac})}  + \frac{c}{6} \left( \frac{\pi^2 \ell^2 }{3 \beta^2 } \gamma^2 (1+v^2) \right) + \mathcal O\left(\frac{1}{\beta^4}\right) 
\end{equation}
This result is exact in $v$ and matches with \cite{Blanco:2013joa} for $d=2$. Note that, from \eqref{bosmlT} it appears that the regulated entanglement entropy $\Delta \mathcal S_A \coloneqq \mathcal S_A -\mathcal S_A^{(\text{vac})}$ diverges 
as $\gamma^2$ when $v \rightarrow 1$. This conclusion, however, is slightly misleading, as far as generic large velocity ($v \rightarrow 1$) behaviour is 
concerned. This is because, if we work out the  $v \rightarrow 1$ limit of \eqref{boostres} at arbitrary temperature, we find 
\begin{equation}\label{exparbT}
\Delta \mathcal S  = \frac{c \ell \pi}{3 \beta} \gamma + \mathcal O \left( \ln \left( 1 - v \right) \right),
\end{equation}
which shows that the leading order divergence in $\Delta \mathcal S$ goes as $\gamma$ for large velocity. Note that the leading order 
behaviour in \eqref{exparbT} is same as that obtained in the large temperature limit, although no such assumption was made to arrive at 
\eqref{exparbT}. This shows that the small temperature result reported for arbitrary dimensions in \cite{Blanco:2013joa}, although valid for arbitrary boost, may not capture the generic large boost behaviour correctly \footnote{Presumably, the coefficient of $\gamma^2$ vanishes when the expression is resummed for arbitrary temperature, and the leading order divergence is of order $\gamma$.}.
\begin{figure}[ht]
\centering
\includegraphics[width = \textwidth]{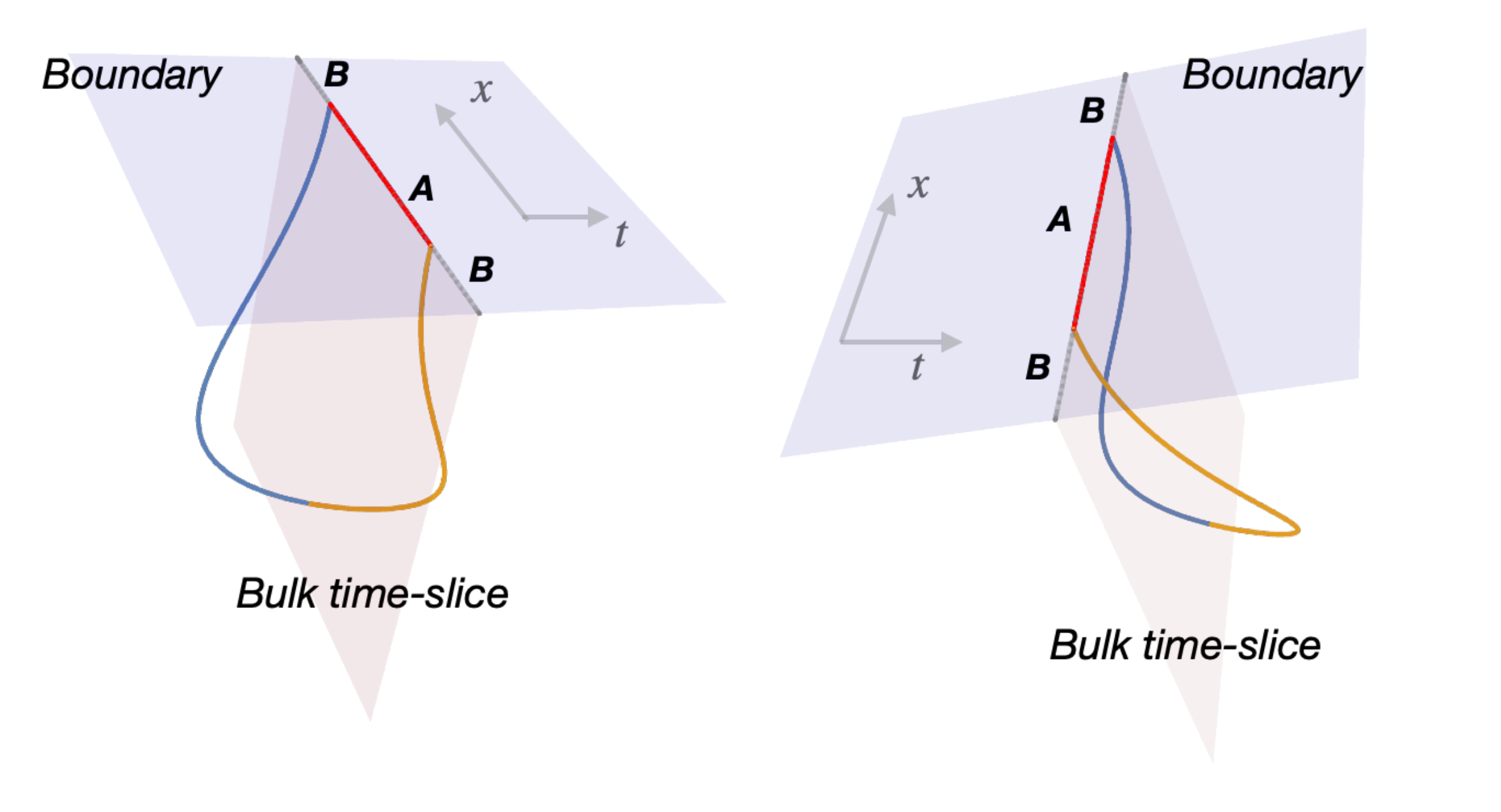}
\caption{Here we present two different views of the same `extremal line' for a subsystem in $d=2$, determined numerically. This plot has been generated for $T=1/2\pi$ and $\ell = 1.12$, when the fluid is moving with a constant vlocity $v = 0.5$. The two solution branches of \eqref{scheomz}, solved with the boundary conditions \eqref{scheombdyz}, have been separately represented by the yellow and the blue lines. The red line denotes the boundary subsystem $A$, while the gray line denotes a canonical time-slice on the boundary. It is noteworthy, that the extremal surface moves out of a constant time-slice in the bulk (denoted by the light-pink surface). }\label{fig:extreme2D}
\end{figure}
%
%
%
%
%
%
\subsubsection*{Numerical verification}\label{sssec:boostbtznum}
%
%
We can adapt the procedure outlined in section \ref{ssec:bbbgen} to d=2, and find a numerical solution to the equations for the extremal surface \eqref{scheom}. 
In fig.\ref{fig:extreme2D}, we provide a representative plot of the extremal surface (which is a line for $d=2$). 
In our solutions we clearly observe that $t(\phi)$ has a non-trivial dependence on the 
world-volume coordinate $\phi$, implying that the extremal surface, 
even in the Schwarzschild coordinates \eqref{schboost}, does not lie on the constant time-slice 
extended into the bulk.

\begin{figure}[thb]
\centering
\includegraphics[width = 0.5\textwidth]{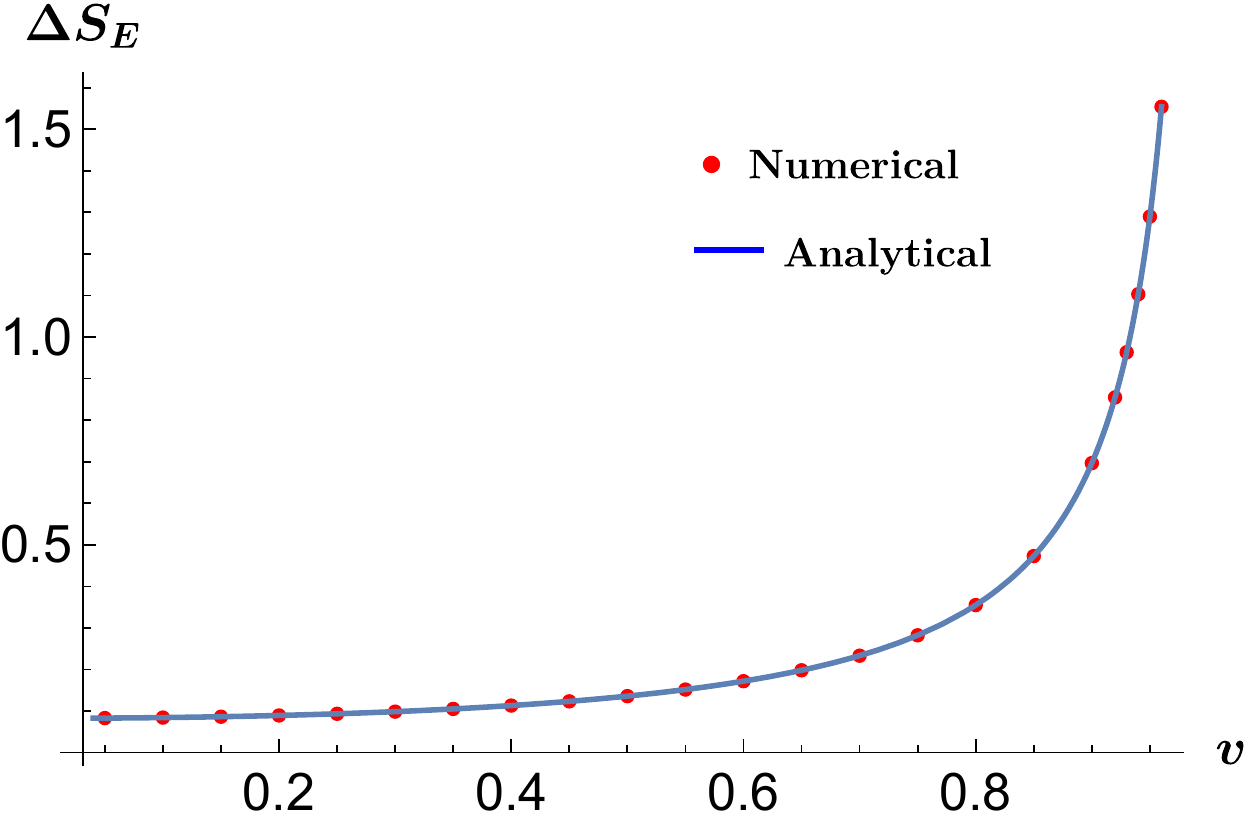}
\caption{Plot of the regulated HEE \eqref{DSdef} for the 1+1D boosted fluid (boosted BTZ) $\Delta S_E$ vs boost $v$. The fixed temperature for this plot is $T = 1/2 \pi$ and subsystem size is $\ell = 2$. Note that since we have regulated our HEE with the vacuum HEE, the $v\rightarrow 0$ limit is non-zero and provides the regulated HEE for the static fluid at fixed temperature.
The answer diverges as $v \rightarrow 1$ following \eqref{exparbT}.}\label{fig:2dnumamatch}
\end{figure}

With this extremal surface determined numerically, we can directly evaluate the regulated HEE for the boosted fluid (bf)
\begin{equation}\label{DSdef}
\Delta \mathcal S =  \mathcal S^{{(\text{bf})}} - \mathcal S^{(\text{vac})}  \equiv \frac{c}{6} ~ \DS
\end{equation}
Here, the pre-factor $c/6 = \mathcal R_{\text{AdS}} / 4 G_N$  has been introduced as a normalization which is convenient for presenting numerical results\footref{foot:Rads}. Our numerical results are exact in boost parameter $v$ and valid for arbitrary temperature. 
In fig.\ref{fig:2dnumamatch}, we numerically plot $\DS$ as a function of $v$ at fixed 
$\ell T$ and we find an excellent match with the analytical answer in \eqref{boostres}. In particular, the 
divergence structure \eqref{exparbT} is correctly reproduced by the numerics. 

%
%

\subsubsection{Holographic Mutual information}\label{sssec:HMIbb2D}
%
%
%
%
%
%
%
Another useful measure of quantum entanglement for a given state is quantum mutual information. 
For non-overlapping subsystems $A$ and $B$, mutual information 
can be expressed in terms of entanglement entropy as follows 
\begin{equation} \label{MIformula}
 I(A,B) = S_A + S_B - S_{A \cup B}, 
\end{equation}
where $S_A$ represents the entanglement entropy between the subsystem $A$ and its compliment. Given 
the property of strong subadditivity of entanglement entropy, this quantity is guaranteed to be positive
($I(A,B) \geq 0$). In addition, it is UV finite. With the help of \eqref{MIformula} the holographic 
prescription for entanglement entropy can be adapted to compute quantum mutual information
(see \cite{Nishioka:2009un,Swingle:2010jz,Fischler:2012uv}). 

For the strip-shaped subsystems, let us consider both $A$ and $B$ to have the same 
size $\ell$, while the distance between the two is denoted by $x$. For the fluid flows 
with a constant velocity $v$, $I(A,B)$  depends on three dimensionless parameters. 
Apart from the fluid velocity $v$, the other two parameters are the ones which may be constituted 
with dimensionless ratios between $\ell$, $x$ and temperature $T$. Since, we have observed 
a peculiar diverging behaviour of entanglement entropy as the fluid velocity approaches 
its relativistic upper bound, we would now like to understand the behaviour of mutual information 
for large fluid velocities. For $d=2$, the existence of the analytical result \eqref{boostres} immediately implies 
an analytical formula for mutual information $I(A,B) = I (\ell, x, \beta, v) $, which is given by 
{ \small{
\begin{equation}\label{MI2dexct}
I (\ell, x, \beta, v) = 
 \frac{c}{6}\ln \left(\dfrac{\sinh^2 \left(\frac{\pi \alpha (v) \ell}{\beta}\right)~\sinh^2 \left(\frac{\pi  \ell}{\alpha (v)\beta}\right)}{\sinh \left(\frac{\pi \alpha (v) x}{\beta}\right)~\sinh \left(\frac{\pi x}{\alpha (v)\beta}\right)~\sinh \left(\frac{\pi \alpha (v) ( 2 \ell +x)}{\beta}\right)~\sinh \left(\frac{\pi (2 \ell+x)}{\alpha (v)\beta}\right)}\right), 
\end{equation}
}}
where $\alpha(v)$ is given in \eqref{xitaus} and $\beta$ is the inverse temperature \eqref{temp}.
When the subsystems $A$ and $B$ are very far away from each other $I(A,B)$ vanishes generically. In fact,
 $I(A,B)$ continuously reduces to zero at some 
finite critical distance ($x_c$) and remains zero for larger distances between $A$ and $B$.
This fact is well understood in the holographic context \cite{Fischler:2012uv}. 
The term $S_{A\cup B}$ in \eqref{MIformula} has two extremal surfaces homologous to $A\cup B$. 
At the critical distance, dominance flips between these two extremal surfaces. 
This critical distance ($x_c$) depends on the properties of the underlying state, for example,  it is determined by the temperature for a thermal state. 
In \cite{Fischler:2012uv}, the functional dependence of this critical distance 
on temperature was computed for $v=0$. Given the exact result \eqref{MI2dexct} in $d=2$, we can immediately generalize their result  
for a non-zero fluid velocity.

\begin{figure}
\centering
\begin{subfigure}{.5\textwidth}
  \centering
  \includegraphics[width=0.9\linewidth]{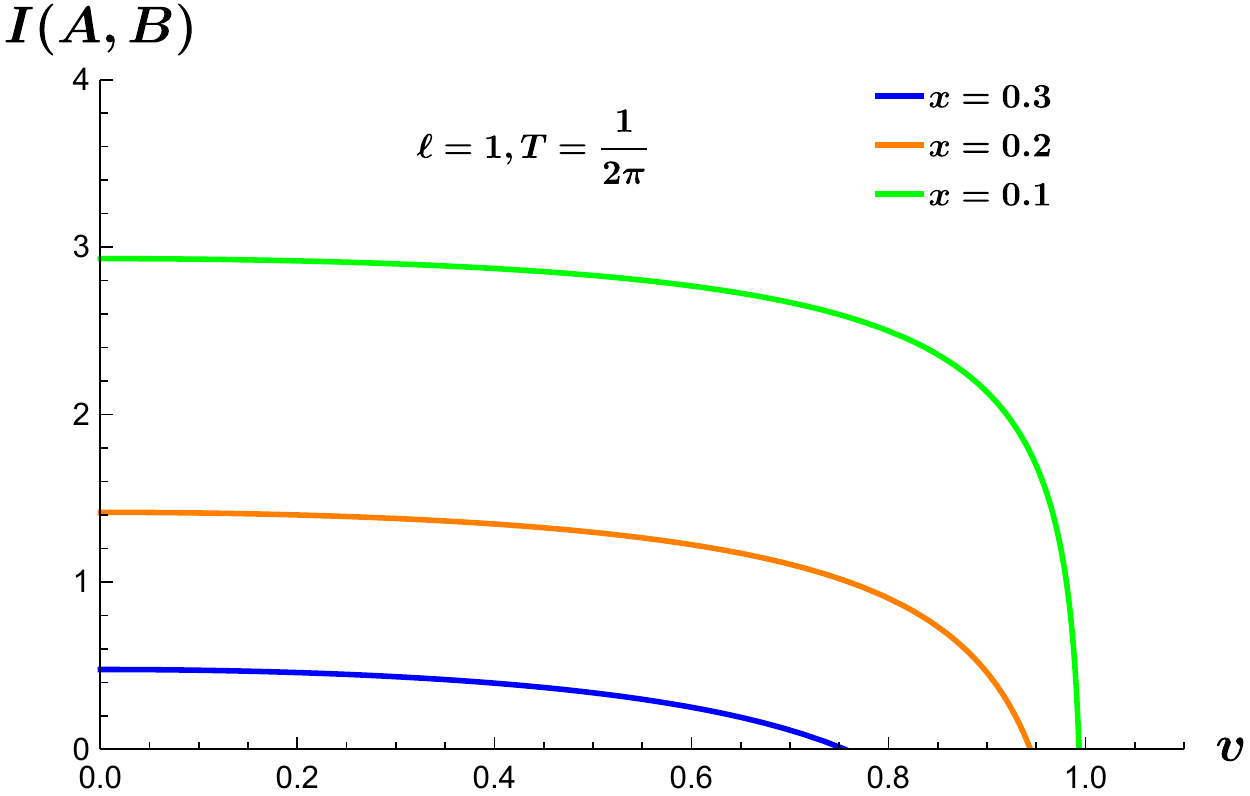}
  \caption{}\label{fig:MIvsv}
\end{subfigure}%
\begin{subfigure}{.5\textwidth}
  \centering
  \includegraphics[width=0.9\linewidth]{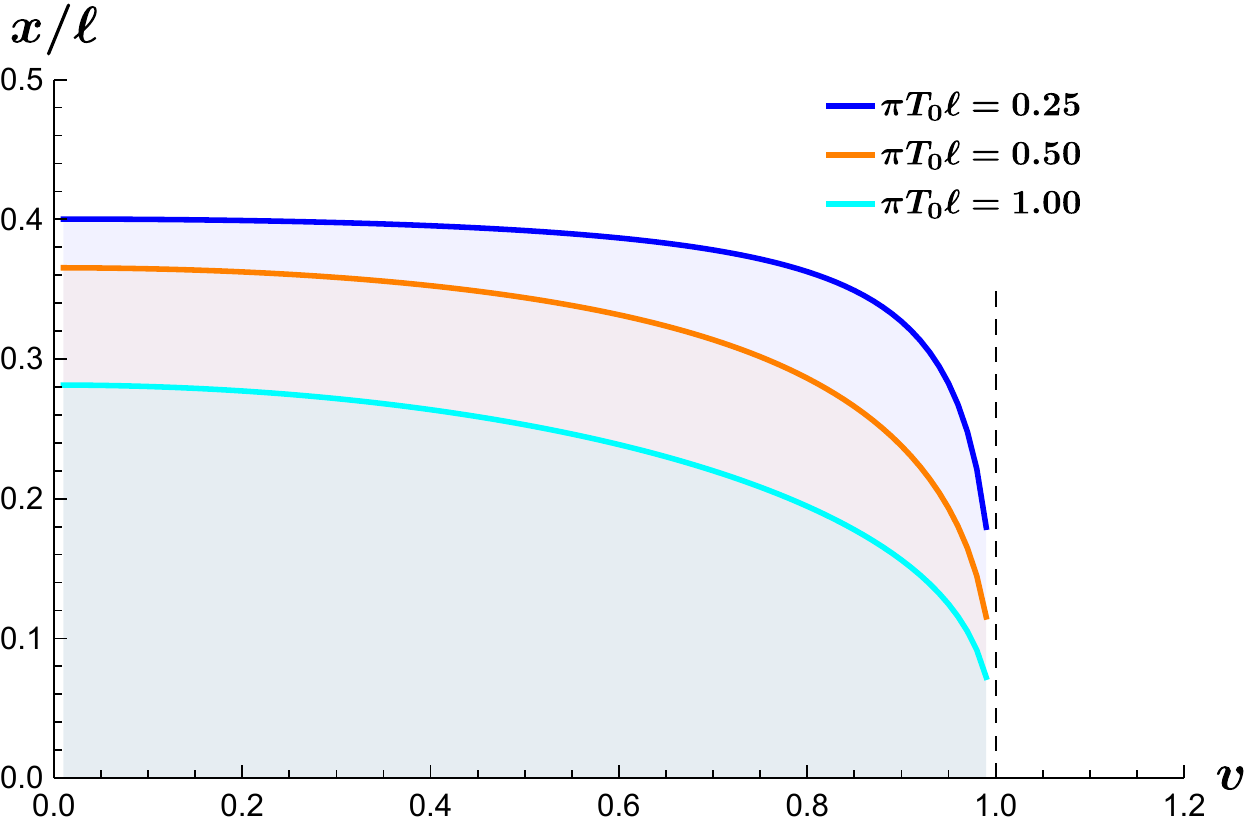}
  \caption{}\label{fig:MI2dxcrit}
\end{subfigure}
\caption{In (a) we plot the mutual information $I(A,B)$, using \eqref{MI2dexct}, as a function of 
the fluid velocity $v$, for fixed $x/\ell$ and $\ell T$. 
Note that there is a critical velocity $v_c$ where $I(A,B)$
vanishes for a given $x/\ell$. We have presented the plots 
corresponding to three different values of separation $x$. Clearly, $v_c$ increases 
with decrease in $x$. In (b), the shaded region below the bold lines,
demarcates the values of $x/\ell$ and $v$ for 
which mutual information is non-zero. This region of non-zero $I(A,B)$
shrinks with the increase in temperature.}
\label{fig:MI2d}
\end{figure}

In fig.\ref{fig:MIvsv}, we have plotted $I(A,B)$ versus the fluid velocity $v$ at fixed $x/\ell$ and
$T \ell$. We observe that, for a given separation 
$x$, there exists a critical velocity 
$v_c$ where mutual information vanishes. The 
value of $v_c$ increases as we decrease $x$,
approaching its relativistic upper bound 
as $x \rightarrow 0$. Once the mutual 
information reaches zero, 
it remains so beyond $v_c$. At $v_c$, there 
is an exchange of dominance between 
the two extremal surfaces for $S_{A \cup B}$. 
This is exactly the same phenomenon 
which gives rise to a critical distance $x_c$ 
between $A$ and $B$, as we have described above. 
Here, we find that the deformation of 
the extremal surfaces, produced due to the fluid 
velocity (see fig.\ref{fig:extreme2D}), forces this exchange of dominance between the extremal surfaces, 
even when the two subsystems are close to each other ($x < x_c$). In fig.\ref{fig:MI2dxcrit} we demarcate the region in the $\{x/\ell,v\}$ plane, where mutual information is non-zero for a fixed temperature. As expected from the results of \cite{Fischler:2012uv}, this region shrinks with the increase in temperature. 
\subsection{Higher dimensional fluids}\label{ssec:bbinD}
%
In higher dimensions, we do not have any analytical results. But, it is straightforward 
to generalize the verified numerical procedure outlined for $d=2$ in section \ref{sssec:HEEbb2D}. 
The general procedure has already been outlined in section \ref{ssec:bbbgen} for general dimensions. 
Therefore, skipping the details, we directly present the results of our numerical computations in this section. 

\begin{figure}[tb]
\centering
\begin{subfigure}{.5\textwidth}
  \centering
  \includegraphics[width=0.9\linewidth]{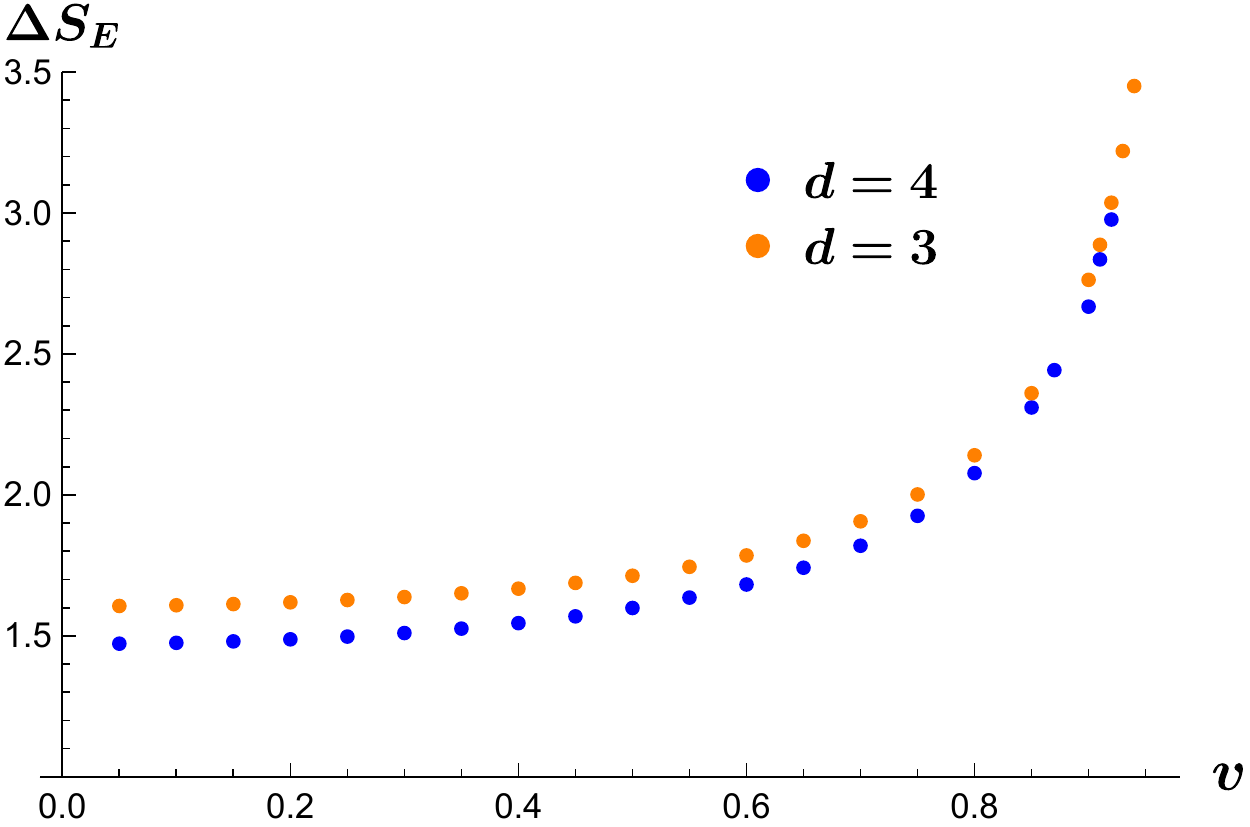}
  \caption{}
  \label{fig:D34delSvsv}
\end{subfigure}%
\begin{subfigure}{.5\textwidth}
  \centering
  \includegraphics[width=0.9\linewidth]{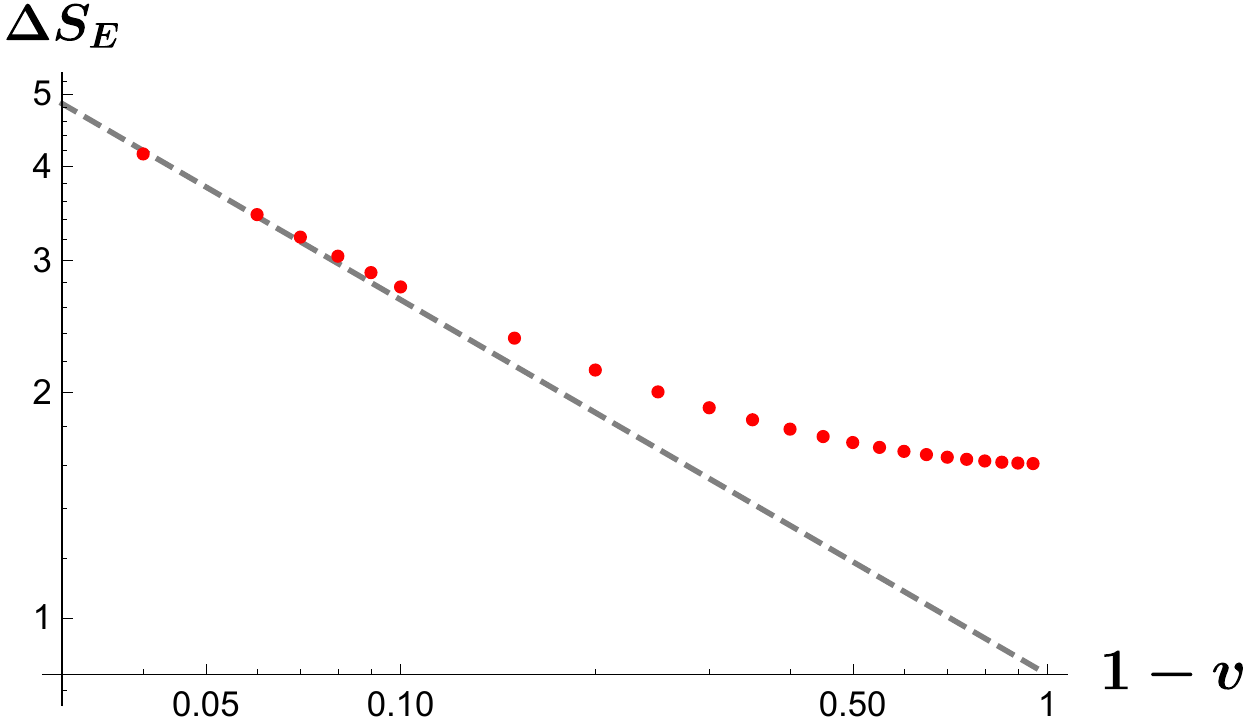}
  \caption{}
  \label{fig:D3div}
\end{subfigure}
\caption{In (a) we plot the variation of regulated entanglement entropy $\Delta S_E$
as a function of the fluid velocity in $d=3,4$ for constant $\ell T = d/4\pi$. 
Note that $\Delta S_E$ diverges as $v \rightarrow 1$. 
In (b) we have displayed a log-log plot 
of $\Delta S_E$ versus $\epsilon_v = (1-v)$ in $d=3$, and for small values of $\epsilon_v$ we observe that the numerically computed  values of $\Delta S_E$ tends towards the dashed straight line with slope $-1/2$. This demonstrates 
that $\Delta S_E$ scales as $(1-v)^{-1/2}$ as $v \rightarrow 1$.}
\label{fig:highDboost}
\end{figure}

The variation of the regulated HEE \eqref{DSdef} versus the fluid velocity $v$, keeping $\ell T$ fixed, 
has been plotted in fig.\ref{fig:D34delSvsv} for $d=3,4$. We observe that just like in $d=2$, the regulated HEE diverges as $v\rightarrow 1$. In fig.\ref{fig:D3div}, we demonstrate that the divergence occurs as $\Delta S_E \sim (1-v)^{-1/2}$ as $v \rightarrow 1$. Again, this behaviour is identical to that observed for $d=2$. With this observation, it is extremely tempting to conjecture that, 
in this set-up, the divergence in $\Delta S_E$ occurs due to a factor of $\gamma$. Perhaps this is universal across dimensions for a strip-shaped subsystems, with the fluid moving parallel to its short edge. 
It would be extremely interesting to understand the nature of this divergence for more general subsystem geometries.  

\begin{figure}[ht]
\centering
\includegraphics[width = 0.7\textwidth]{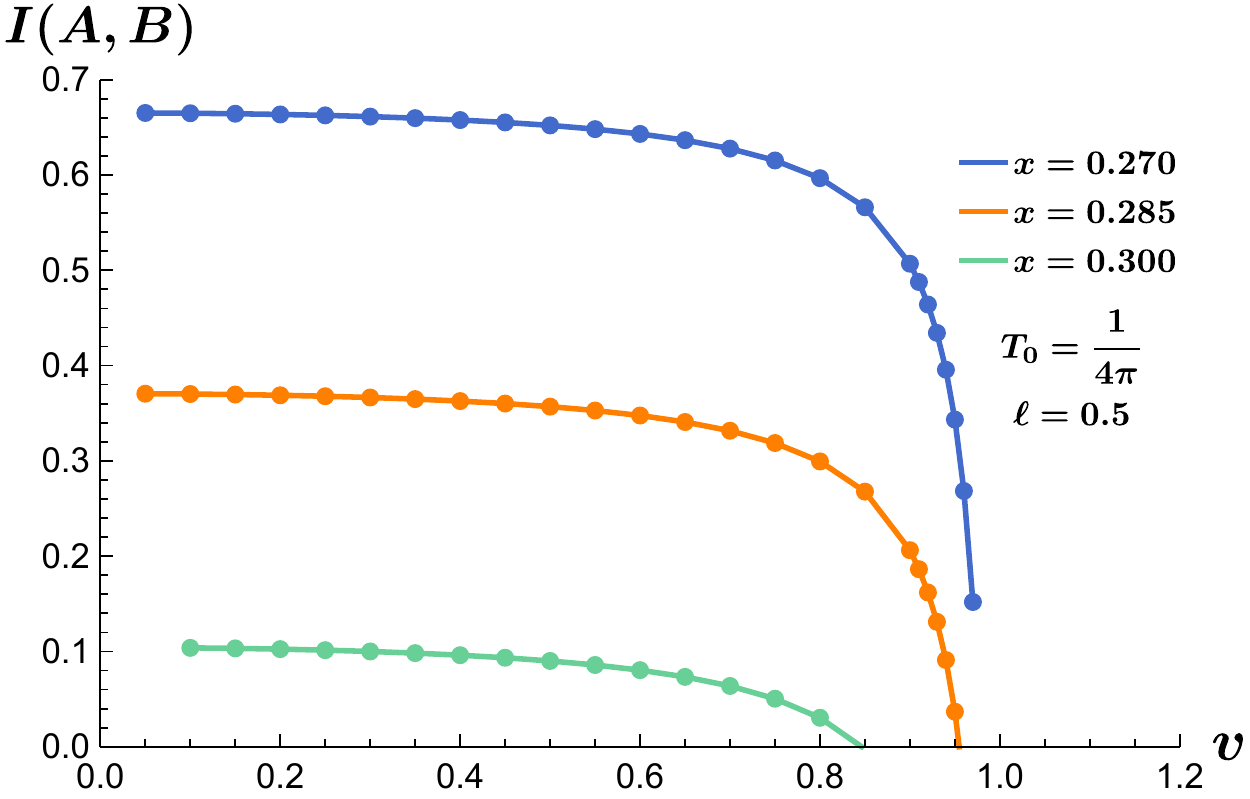}
\caption{Numerical plots exhibiting the variation of mutual 
information $I(A,B)$ with respect to fluid velocity $v$ in $d=3$.}\label{fig:MIhighD}
\end{figure}

Once we are able to compute the entanglement entropy with reliable numerics, it is straightforward to 
compute the mutual information \eqref{MIformula} in higher dimensions. In fig.\ref{fig:MIhighD}, 
we plot the mutual information versus the fluid velocity in $d=3$. 
Even here, we notice a behaviour which is 
qualitatively identical to that in $d=2$ (see fig.\ref{fig:MIvsv}). In particular, we observe that 
there is a critical velocity beyond which the mutual 
information vanishes. Again, this feature is also perhaps true across all dimensions. 

%
%
%
%
%
%
%
%
%
%
%
%
%
%
%
%
%
%
%
%
%
%
%
%
%
%
%
%
%
%
%
%
%
%
\section{Entanglement in hydrodynamic sound mode}\label{sec:hydromodes}
%
\subsection{Outline of the general procedure}\label{ssec:genpro}
%
After considering the steady state fluid configurations with a constant velocity, 
in this section, we proceed to analyse the entanglement structure of 
fluid states which are slightly out of equilibrium. More specifically, we shall focus on the propagating sound mode, 
which is ubiquitous in all systems admitting a macroscopic fluid description. Our set-up here, will again be holographic, 
which technically limits our observations to conformal relativistic fluids. However, our qualitative observations 
in this section would perhaps be more universal than the holographic context. 

The sound mode is a fluctuating small amplitude perturbation over a
stationary background\footnote{For simplicity, throughout our analysis in this section, we shall consider the sound mode 
over a static background. Our analysis can be straightforwardly generalized to the case with a background fluid velocity (see 
section \ref{sec:disco} for a discussion about this point). \label{foot:static}}. We will compute the HEE for the sound mode using the fluid-gravity 
correspondence \cite{Bhattacharyya:2007vjd}. We shall now provide a brief overview of the general procedure adopted for this HEE computation. 

The dual black brane solution providing a holographic description of the fluctuating states of a relativistic conformal fluid in $d=4$, with a controlled derivative expansion, was first provided in \cite{Bhattacharyya:2007vjd}, which was subsequently generalized to other dimensions in \cite{VanRaamsdonk:2008fp,Haack:2008cp,Bhattacharyya:2008mz}. Using these constructions, it is simple to 
read off the black brane geometries dual to the sound modes.  Since these bulk geometries are constructed 
in the derivative expansion, they are valid when the wavelength ($\lambda = 2 \pi / k$) of the sound mode 
is large compared to the temperature scale ($\lambda T \gg 1$). Beside this small parameter $k/T$, the amplitude of 
the sound mode (say, $\epsilon$) also must be small\footnote{As we will see in the next subsection, $\epsilon$ is the dimensionless 
parameter which may be taken as the ratio $\delta T / T_0$, where $\delta T$ is the amplitude of temperature fluctuations induced 
by the sound mode, while $T_0$ is the background temperature.}, since it is a linearized solution to the fluid equations. 
We shall consider the amplitude ($\epsilon$) as the smallest of the two parameters.

Due to the linearized nature of the sound mode, the dual geometry of the fluctuating black brane has the following structure 
\begin{equation} \label{gengexp}
ds^2_{d+1} = g_{ab} ~dx^a dx^b = \left( g^{(0)}_{ab} + \epsilon ~g^{(1)}_{ab} \right) dx^a dx^b
\end{equation}
Here $g^{(1)}_{ab}$ is the $\epsilon$-order correction to the background metric $g^{(0)}_{ab}$, and we shall consider $g^{(0)}_{ab}$ 
to be the static black brane dual to the static fluid\footref{foot:static}. 
Since the background is static there is no $k$ dependence in $g^{(0)}_{ab}$, while $g^{(1)}_{ab}$ has a further expansion 
in powers of $k$ due to the derivative expansion corrections to the bulk metric. 

Similar to the previous section, we would like to compute HEE for a strip-like subsystem over the fluid configuration
with a propagating sound mode (see fig.\ref{fig:Scheme}). The short-edge of the strip is of length $\ell$, and again, to keep 
the computations simple, we shall restrict ourselves to the scenario where the sound mode is propagating along the direction of the short edge\footnote{This can be easily generalized to the case where there is an angle between the direction of the short edge and that of sound propagation. Naively, it may be expected that the component of the sound mode 
along the long-edge of the strip would not contribute
to $\Delta S_E$.}.
Apart from $k/T$ and $\epsilon$, the dimension $\ell$ of the subsystem 
introduces a third independent dimensionless parameter in the problem, which is given by
$T \ell$. While there are no a priori restrictions on this parameter, in the holographic set-up, if we are to capture 
the thermal effects adequately, we must ensure that $T \ell \sim \mathcal O (1)$ or greater.  

The HEE computation is accomplished by computing the area of the extremal surface 
as prescribed in \eqref{EntFormula} \cite{Hubeny:2007xt}. Unlike the stationary scenario discussed in section  \ref{sec:bbb}, we now expect 
a non-trivial time dependence of HEE. If we parameterize our co-dimension 2 extremal surface ($\Gamma$) with  the coordinates $\{\xi\}$, 
the induced metric on $\Gamma$ takes the form 
\begin{equation}\label{indmetsch}
 \mathcal G_{\alpha \beta} = \frac{\partial x^a}{\partial \xi^\alpha} \frac{\partial x^b}{\partial \xi^\beta} g_{ab}
\end{equation}
However, due to the decomposition \eqref{gengexp}, the induced metric also decomposes as 
\begin{equation}\label{indmetdec}
 \mathcal G_{\alpha \beta} = \mathcal G^{(0)}_{\alpha \beta} + \epsilon ~ \mathcal G^{(1)}_{\alpha \beta},
\end{equation}
where
\begin{equation}
\mathcal G^{(0)}_{\alpha \beta} = \frac{\partial x^a}{\partial \xi^\alpha} \frac{\partial x^b}{\xi^\beta} g^{(0)}_{ab} , ~~\mathcal G^{(1)}_{\alpha \beta} = \frac{\partial x^a}{\partial \xi^\alpha} \frac{\partial x^b}{\xi^\beta} g^{(1)}_{ab}. 
\end{equation}
The area of the co-dimension 2 surface $\Gamma$ is immediately given by the induced metric 
\begin{equation}
 A_{\Gamma} = \int \left(d\xi\right)^{d-1} \sqrt{\mathcal G}
\end{equation}
where $\mathcal G$ is the determinant of the induced metric \eqref{indmetsch}. However, due to 
\eqref{indmetdec} this area also has an expansion in amplitude 
\begin{equation}
 A_{\Gamma} = A^{(0)}_{\Gamma} + \epsilon ~A^{(1)}_{\Gamma}
\end{equation}
Here $A^{(0)}_{\Gamma}$ has the contribution from the background static geometry and constitutes 
the leading order piece in HEE. It is given by 
\begin{equation} \label{HEEeq}
A^{(0)}_{\Gamma} = \int (d\xi)^{d-1} \sqrt{\mathcal G^{(0)}}. 
\end{equation} 
By the prescription \eqref{EntFormula}, this area when evaluated over the extremal surface provides the entanglement of the strip subsystem for the fluid in static equilibrium. On the other hand the corrections due to the sound mode is captured by 
$A^{(1)}_{\Gamma}$, which is given by 
\begin{equation} \label{genformu}
A^{(1)}_{\Gamma} = \frac{1}{2}  \int (d\xi)^{d-1} \sqrt{\mathcal G^{(0)}} ~\mathcal G^{(1)}_{\alpha \beta} {\mathcal G^{(0)}}^{\alpha \beta}. 
\end{equation}
Note that, at first order in $\epsilon$, the functional in \eqref{genformu} is to be evaluated over the same surface $\Gamma$ that was obtained by extremizing $A^{(0)}$ in \eqref{HEEeq}. The change in HEE produced due to the fluid sound mode is therefore given by
\begin{equation}\label{deltaSsm}
 \Delta \mathcal S = \frac{A^{(1)}_{\Gamma}}{4 G_N} 
\end{equation}
As in the previous section (see \eqref{Sedefb}), for our numerical plots, it will be 
convenient to work with the quantity $\DS =  A^{(1)}_{\Gamma}/ L^{(d-2)}$.

\subsection{Sound mode in $d=2$}\label{sound2D}

In $d=2$, conformal relativistic fluid dynamics is trivial and 
there are no non-trivial transport coefficients, in the uncharged sector\footnote{For example, in $d=2$, at first order in derivative expansion, due to the low number of dimensions, there 
is no sheer, while a bulk viscosity is not allowed by conformal invariance. If there are additional conserved 
currents, then in $d=2$ there can be dissipative transport, such as diffusion of the associated conserved charge.}. 
Therefore, the energy-momentum tensor is always of the perfect fluid form 
\begin{equation}\label{emt2d}
 T_{\mu \nu} = \mathcal E (T) u_{\mu} u_{\nu} + \mathcal P (T) P_{\mu \nu}. 
\end{equation}
The key difference of \eqref{emt2d} with that of in higher dimensions \eqref{steadyperflu}, is that, \eqref{emt2d} does not 
have any higher derivative corrections, even if the fluid variables ($T$ and $u^\mu$) have arbitrary space-time dependence. 
Here, the energy density $\mathcal E$ and pressure $\mathcal P$ must be equal (see \eqref{PEperflu}) so as to ensure tracelessness of the energy-momentum tensor - a requirement of conformality. All solutions to this system are non-dissipative. 
The two fluid variables - $T$ and the spatial component of normalized $u_\mu$ - are to be determined by the fluid equations 
which are given by the conservation of the energy-momentum tensor
\begin{equation}\label{conemt2d}
 \partial_{\mu} T^{\mu \nu} = 0 
\end{equation}
Here, \eqref{conemt2d} gives us two independent equations, providing solutions for the two fluid variables. These solutions 
can have a non-trivial dependence on space-time coordinates, without requiring higher derivative corrections. Thus, unlike higher dimensions, the restrictions due to derivative expansion on the sound wave 
momentum $k$ (i.e. $k/T \ll 1$ ) does not apply for the $d=2$ fluid. 

For any arbitrary solution of \eqref{conemt2d}, an explicit holographic dual metric was constructed in \cite{Haack:2008cp} 
\begin{equation}\label{gravmet2d}
\begin{split}
 ds^2 &= \frac{1}{z^2} \left( - \left( 1- \frac{z^2}{\zh^2} \right) u_\mu u_\nu dx^\mu dx^\nu + P_{\mu \nu} dx^\mu dx^\nu 
 + 2 u_\mu dx^\mu dz \right)\\
 & + \frac{2}{z} \partial_\lambda u^\lambda ~u_\mu u_\nu dx^\mu dx^\nu 
 - \frac{1}{z} u^\lambda \partial_\lambda \left( u_\mu u_\nu \right) dx^\mu dx^\nu
\end{split}
\end{equation}
The derivative corrections in the second line of \eqref{gravmet2d} ensures that the metric is an exact solution
to the Einstein equations, whenever the fluid parameters $u^\mu(x)$ and $T(x)= 1 / \left( 2 \pi \zh(x) \right)$, 
as functions of the boundary coordinates $x^\mu$, solves \eqref{conemt2d}. The explicit form of the fluid equations \eqref{conemt2d} is given by
\begin{equation}\label{2dflueq}
 u^\mu \partial_\mu \zh = \zh \partial_\nu u^\nu, ~~P_\mu^{~\nu} \partial_\nu \zh = \zh u^\lambda \partial_\lambda u_\mu.  
\end{equation}

The metric \eqref{gravmet2d} is essentially diffeomorphic to the BTZ black hole \cite{Haack:2008cp}, which on the gravity side, reflects the triviality of conformal hydrodynamics in $d=2$. The dynamics induced by these non-trivial diffeomorphisms serves as a toy model for hydrodynamics with a non-dissipative sound mode. Note that, \eqref{gravmet2d} is similar to a fluid-gravity metric in higher dimensions (see \eqref{gravmet3d}, \eqref{gravmet4d}) expanded upto first order in derivatives, but \eqref{gravmet2d} is an exact solution. This absence of the requirement of a derivative expansion makes this setting 
particularly suitable to initiate a study of the HEE for the sound mode. 
\subsubsection*{Linearized solutions}
Let us denote the boundary coordinates with $x^\mu = \{ t, x\}$ 
and express the two fluid functions in the following way 
\begin{equation}
 u^\mu = \gamma \{ - 1 , v \}, ~~ T = \frac{1}{2\pi \zh}, ~~\text{where}~ \gamma = \frac{1}{\sqrt{1-v^2}} 
\end{equation}
Then \eqref{2dflueq} admits the following linearized sound wave solution 
\begin{equation}\label{2dsinsol}
 \begin{split}
  \zh &= \zh_0  \left( 1 + \epsilon  \cos \left( \omega t - k x\right) + \mathcal O \left( \epsilon^2 \right)\right) ~,~~\\
  v &= \epsilon \cos \left( \omega t - k x\right) + \mathcal O \left( \epsilon^2 \right), ~~ \text{with} ~ \omega = k,   
 \end{split}
\end{equation}
where $\zh_0$ determines the background temperature. 

Exploiting linearization, it is straightforward to superimpose these sinusoidal solutions with a Gaussian weight to constitute a propagating wave-packet solution of the form 
\begin{equation}\label{2dpulsol}
 \begin{split}
  \zh &= \zh_0  \left[ 1 - \epsilon ~ \exp \left( - \frac{\left( x- t\right)^2 }{2 \sigma^2}\right)  + \mathcal O \left( \epsilon^2 \right)\right] ~,~~\\
  v &= \epsilon ~ \exp \left( - \frac{\left( x- t\right)^2 }{2 \sigma^2}\right) + \mathcal O \left( \epsilon^2 \right),
 \end{split}
\end{equation}
where $\sigma$ is the width of the pulse. Clearly, \eqref{2dpulsol} is also a linearized solution of the fluid equations \eqref{2dflueq}. Note that, physically this solution corresponds to a energy density or pressure pulse propagating in the positive $x$-direction. The amplitude of this pulse is proportional to $\epsilon$, which in general, denotes a small positive or negative correction to their background values. In our subsequent numerical computations, we shall consider $\epsilon$ in \eqref{2dpulsol} to be positive, corresponding to a positive correction to the background energy density. 

We can plug in the solutions \eqref{2dsinsol} and \eqref{2dpulsol} into \eqref{gravmet2d} to obtain the corresponding bulk metric, which will be clearly of 
the form \eqref{gengexp}. We can then use the procedure outlined in section \ref{ssec:genpro} to compute corrections to HEE corresponding to these 
fluid solutions. 

\subsubsection*{Extremal surface in the background}

In order to implement the algorithm of section \ref{ssec:genpro}, we need to find the extremal surface in the background metric (see \eqref{gengexp})
\begin{equation}\label{metbtz}
 g_{a b}^{(0)} dx^a dx^b = \frac{1}{z^2} \left( - \left( 1 - \frac{z^2}{\zh_0^2} \right) dt^2 - 2 dt dz + dx^2 \right) 
\end{equation}
This is simply the static BTZ black brane written in Eddington-Finkelstein (EF) coordinates and is obtained by plugging in the $\mathcal O (\epsilon^0)$ terms 
of \eqref{2dsinsol} and \eqref{2dpulsol} into \eqref{gravmet2d}.  

If we now consider a boundary subsystem ($A$) of length $\ell$ centered at the origin ($x \in [- \ell/2 , \ell/2]$). 
Here, the extremal surface in EF coordinates anchored on and homologous to $A$ can be written down analytically. It is simply a coordinate transformation (see appendix \ref{app:EFcoord}) 
of the corresponding extremal surface in Schwarzschild coordinates \cite{Bhattacharya:2013bna}.
The explicit form of the embedding equations for this extremal surface in EF coordinates is given by 
\begin{equation}\label{2dextsur}
\begin{split}
 & z(\phi) = z_\star ~\cos \phi \\
 & x(\phi) = \zh_0 ~\ln \left( \dfrac{\kappa \sin \phi + \sqrt{1-\kappa^2 \cos^2\phi} }{\sqrt{1 - \kappa^2}} \right) \\
 & t(\phi) = t_b - \text{arctanh} \left( \kappa \cos \phi \right) \\
 & \text{where}~ z_\star = \zh_0 ~\kappa , ~ \text{and} ~ \kappa = \tanh \left( \frac{\ell}{2\zh_0} \right) . 
\end{split} 
\end{equation}
Here $t_b$ denotes the boundary canonical time-slice which contains the subsystem $\mathcal A$. The dependence of
HEE on $t_b$ provides its time evolution. 

\subsubsection*{Numerical results for HEE}

In order to find $\Delta S_E$, we must evaluate the functional 
\eqref{genformu} over the extremal surface \eqref{2dextsur}. The metric correction $g^{(1)}_{ab}$ in \eqref{gengexp}, and hence the corrections to the induced metric $\mathcal G^{(1)}_{\alpha \beta}$  is obtained by computing the $\mathcal O (\epsilon)$ term from \eqref{gravmet2d} after substituting the linearized solutions \eqref{2dsinsol} and \eqref{2dpulsol}
\footnote{Note that the expression for $\mathcal G^{(1)}_{\alpha \beta}$ is 
significantly long. Therefore, to avoid clutter we refrain from reporting it in this paper. 
It may be  straightforwardly computed with any software supporting symbolic manipulation ( `{\it{Mathematica}}' has been used in our work). The final answer for $\Delta S_E$ is obtained after integrating 
the functional in \eqref{genformu}, which we were only able to perform numerically. }. 
We will now present our results for $\Delta S_E$ in terms of numerical plots.
\begin{figure}[thb]
\centering
\includegraphics[width = 0.5\textwidth]{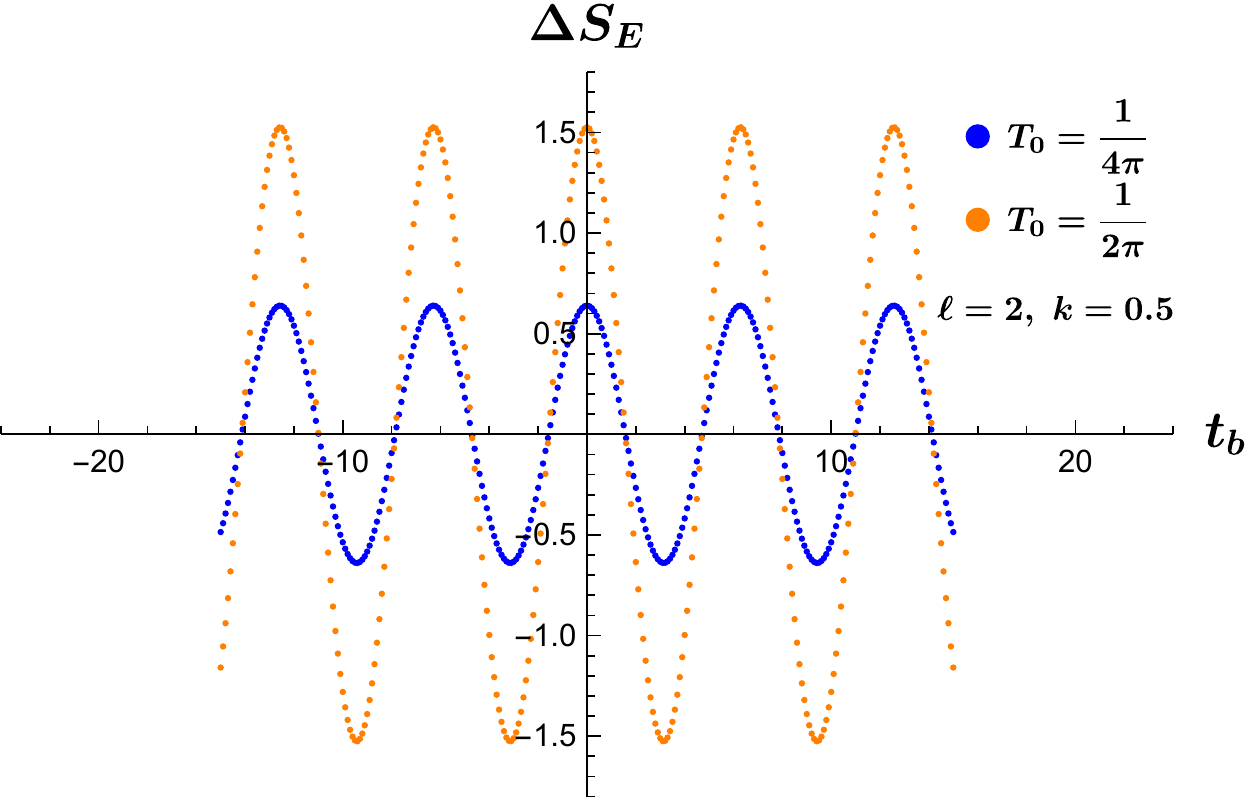}
\caption{$\Delta S_E$ vs $t_b$ for sound wave in $d=2$.}\label{fig:delSswvst}
\end{figure}
\begin{flushleft}
{\it{Sound-waves:}}
\end{flushleft}

We now compute $\DS$ for the scenario where a non-dissipative sound-wave \eqref{2dsinsol} with momentum $k$ constantly passes over 
our subsystem $A$. Once we assume a small value of the parameter $\epsilon$,
there are two more independent dimensionless parameters in the problem - $\ell T_0$ , $k/T_0$, $T_0$ being the background temperature. 

The variation of $\DS$ with boundary-time ($t_b$) has been shown in fig.\ref{fig:delSswvst}. We find a periodic variation of HEE,
which was expected due to the sinusoidal nature of the fluid configuration. The period of the oscillations 
are given by $2 \pi / k$ and is independent of temperature. This was expected from the fact that the time-dependence only arises through 
the term $\cos k t$ appearing in \eqref{2dsinsol}. In the embedding equations \eqref{2dextsur}, $t$ goes linearly with $t_b$, which gives 
rise to a term proportional to $\cos k t_b$ in final linearized answer. 

The amplitude of these oscillations 
is determined both by the temperature and $k$, for a fixed $\ell$. 
From fig.\ref{fig:delSswvst} we see that for fixed $k$, this amplitude increases with increase in temperature.  
Recall that there exists no restriction on the value of $k$ here, due to the lack of any derivative expansion 
for the $d=2$ conformal fluid. So, we can plot this amplitude for arbitrary $k$ at a fixed temperature.
\begin{figure}[thb]
\centering
\begin{subfigure}{.5\textwidth}
  \centering
  \includegraphics[width=0.9\linewidth]{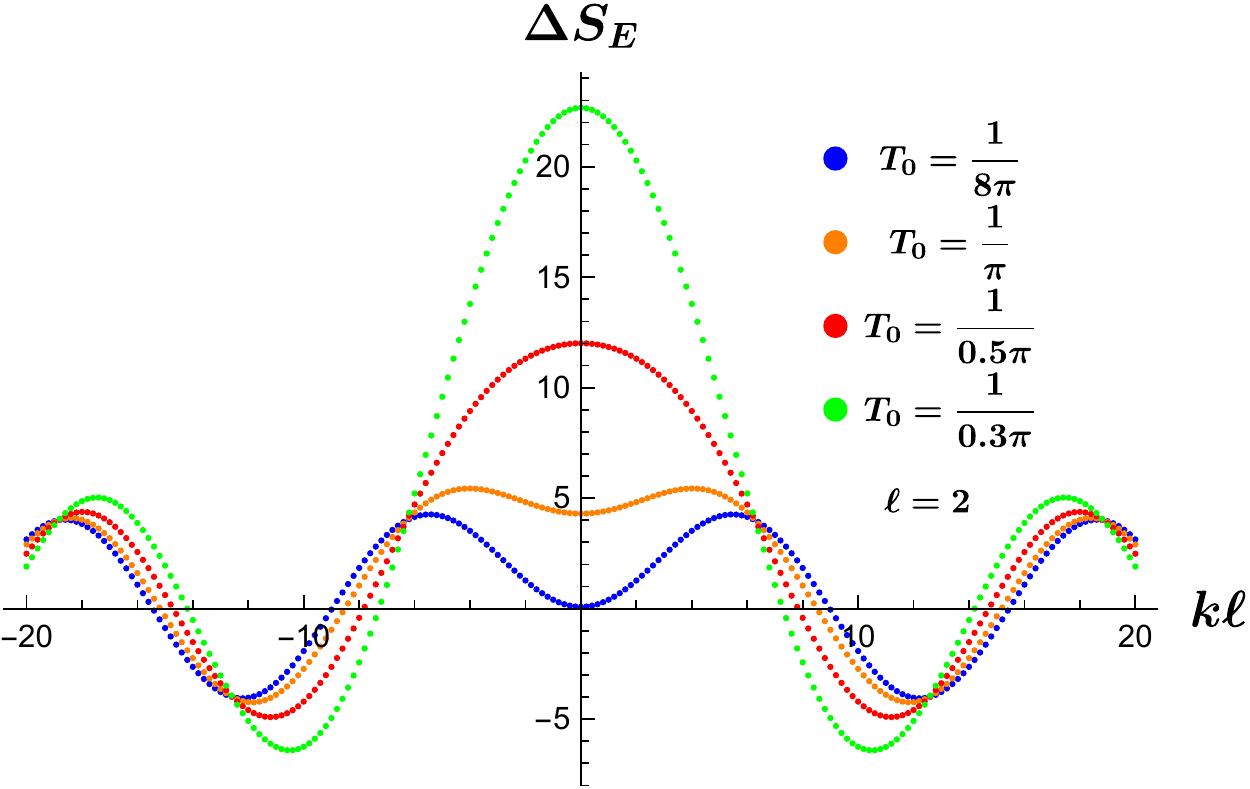}
  \caption{}
  \label{fig:delsvsksw}
\end{subfigure}%
\begin{subfigure}{.5\textwidth}
  \centering
  \includegraphics[width=\linewidth]{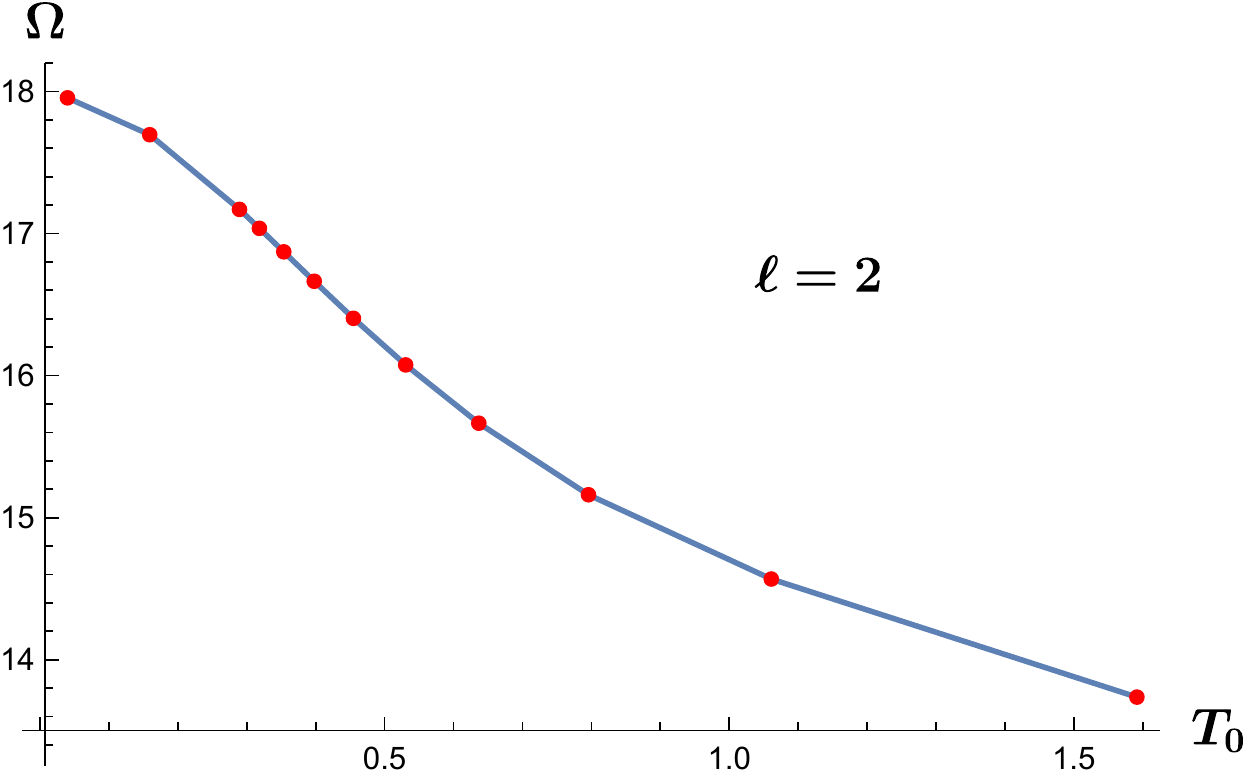}
\caption{}
  \label{fig:OmegavsT}
\end{subfigure}
\caption{In (a) we plot the variation of $\Delta S_E$ at $t_b =0$ as a function of $k \ell$ for sound mode in $d=2$.
This explicates the dependence of the amplitude of $\Delta S_E$ on $k$. We find an interesting quasi-periodic behaviour of 
$\Delta S_E$ with respect to $k$. In (b) we show that the distance between the first two zeros of  $\Delta S_E (t_b=0)$ in (a), 
decreases with increase in temperature.}
\label{fig:test}
\end{figure}
In fig.\ref{fig:delsvsksw}, we study the variation of $\Delta S_E$ with respect to $k$ on a fixed boundary-time slice $t_b =0$.
It is very interesting to note that for large values of $k$, $\Delta S_E$ behave sinusoidally with $k$. In particular, there 
exists special values of $k$ for which $\Delta S_E$ vanishes. The distance $\Omega$ between 
the first two zeros of $\Delta S_E (t_b=0)$, in fig.\ref{fig:delsvsksw}, as well as the period 
of oscillation for large $k\ell$, is determined 
by the temperature. In fig.\ref{fig:OmegavsT}, we have shown that the value of $\Omega$ decreases with the increase in temperature. 
\begin{figure}[tbh]
\centering
\begin{subfigure}{.5\textwidth}
  \centering
  \includegraphics[width=0.9\linewidth]{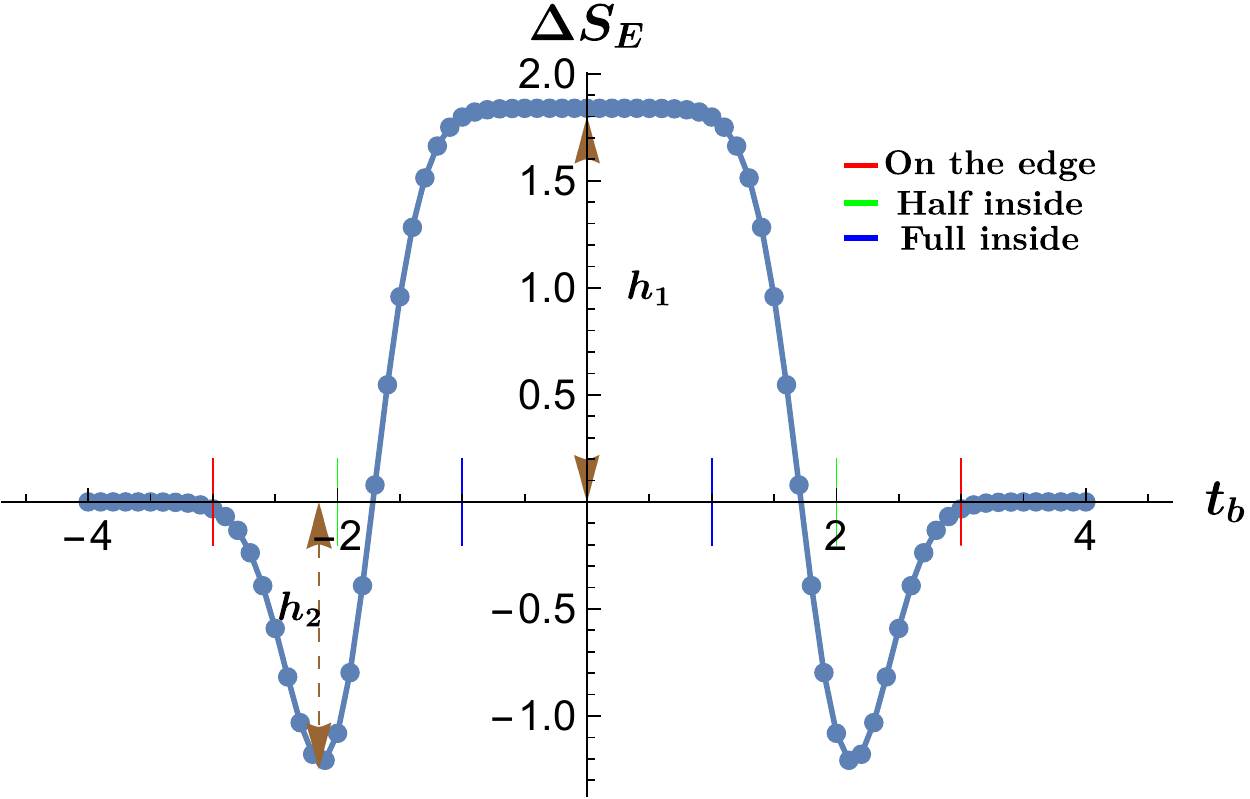}
 \caption{}
  \label{fig:delSpvst1}
\end{subfigure}%
\begin{subfigure}{.5\textwidth}
  \centering
  \includegraphics[width=0.9\linewidth]{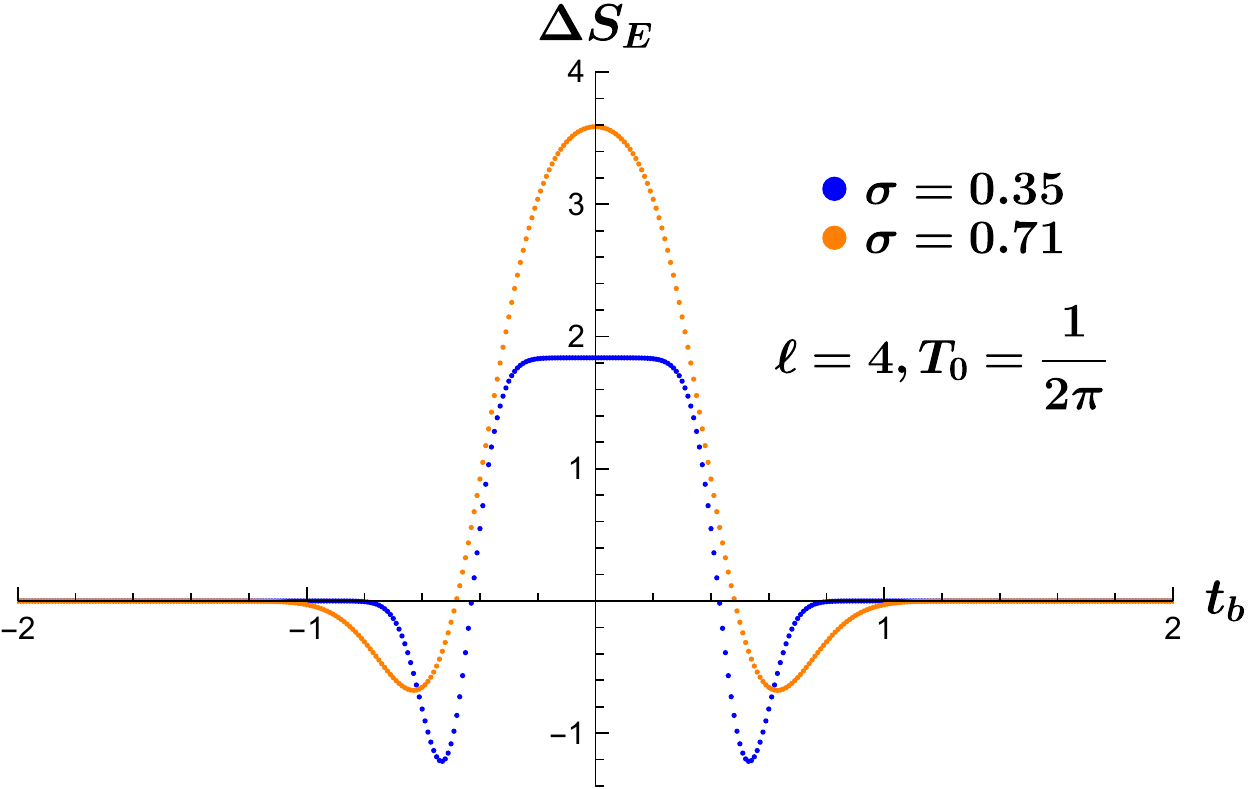}
 \caption{}
  \label{fig:delSpvst2}
\end{subfigure}
\caption{In (a) we plot the variation of $\Delta S_E$ with respect to boundary time $t_b$,
as a pressure or energy density pulse travels across the subsystem $A$. We have 
chosen the pulse width $\sigma = 0.35$, while the size of the subsystem has been taken to be 
$\ell = 4$. The background temperature is kept at $T_0 = 1/2 \pi$. At time $t_b=\mp 3$ (shown 
with the red line), the edge of the pulse touches the subsystem as it starts entering it 
(or completely leaves it).  At time $t_b=\mp 2$, half of the pulse has entered (left) 
the subsystem, which appears to be close to the maximum value of the dip. Finally, at 
$t_b = \mp 1$ the pulse has completely entered the subsystem (or about to leave it). 
Note that, for time-interval ($t_b = -1$ to $t_b=+1$) during which the pulse lives inside the subsystem, $\Delta S_E$ saturates to a
constant value. In (b) we demonstrate how the time evolution of $\Delta S_E$ varies as we change 
the width of the pulse. 
}
\label{fig:delSpvst}
\end{figure}
\begin{flushleft}
{\it{Propagating pulse:}}
\end{flushleft}
We now consider a traveling wave-packet  - 
a pressure or energy density pulse \eqref{2dpulsol}.  
We record the change in HEE as this pulse traverses the subsystem $A$. 
As expected, when the pulse is completely outside the subsystem $A$, travelling towards 
or away from it, $\Delta S_E$ is zero. 
But, while the pulse moves across the subsystem, $\Delta S_E$ exhibits an interesting time dependence, as shown in fig.\ref{fig:delSpvst}. 

After fixing $\epsilon$ to a small value, there are three length scales in this problem as well. Apart from $\ell$ and $T$, there is $\sigma$ which is the width of the pulse. When the pulse of positive energy density moves completely inside the subsystem $A$, it perhaps drags in some `matter' within $A$ from outside. From this naive picture we can expect 
that there will be an increase in $\Delta S_E$, which would eventually go down to zero, when the pulse passes out. In fig.\ref{fig:delSpvst1} we see that 
this expectation is almost correct, except that there 
is a additional feature. When the pulse enters 
(or exits) the subsystem, we find that there is 
an interesting dip in $\Delta S_E$ before 
the expected rise (or fall). 

In fig.\ref{fig:delSpvst2}, we demonstrate 
that if the width of the pulse $\sigma$ is 
comparable to or larger than the size of 
the subsystem $\ell$, there is a unique peak of
$\Delta S_E$, which corresponds to the time 
when the maxima of the peak coincides with 
the central point of the subsystem. On the other 
hand, when $\sigma$ is significantly smaller 
than $\ell$, the $\Delta S_E$ saturates 
to a constant value for the duration 
the pulse spends inside the subsystem. 
In both the case, the dip in $\Delta S_E$,
during entry or exit, persists. 

\begin{figure}[tbh]
\centering
\begin{subfigure}{.5\textwidth}
  \centering
  \includegraphics[width=0.9\linewidth]{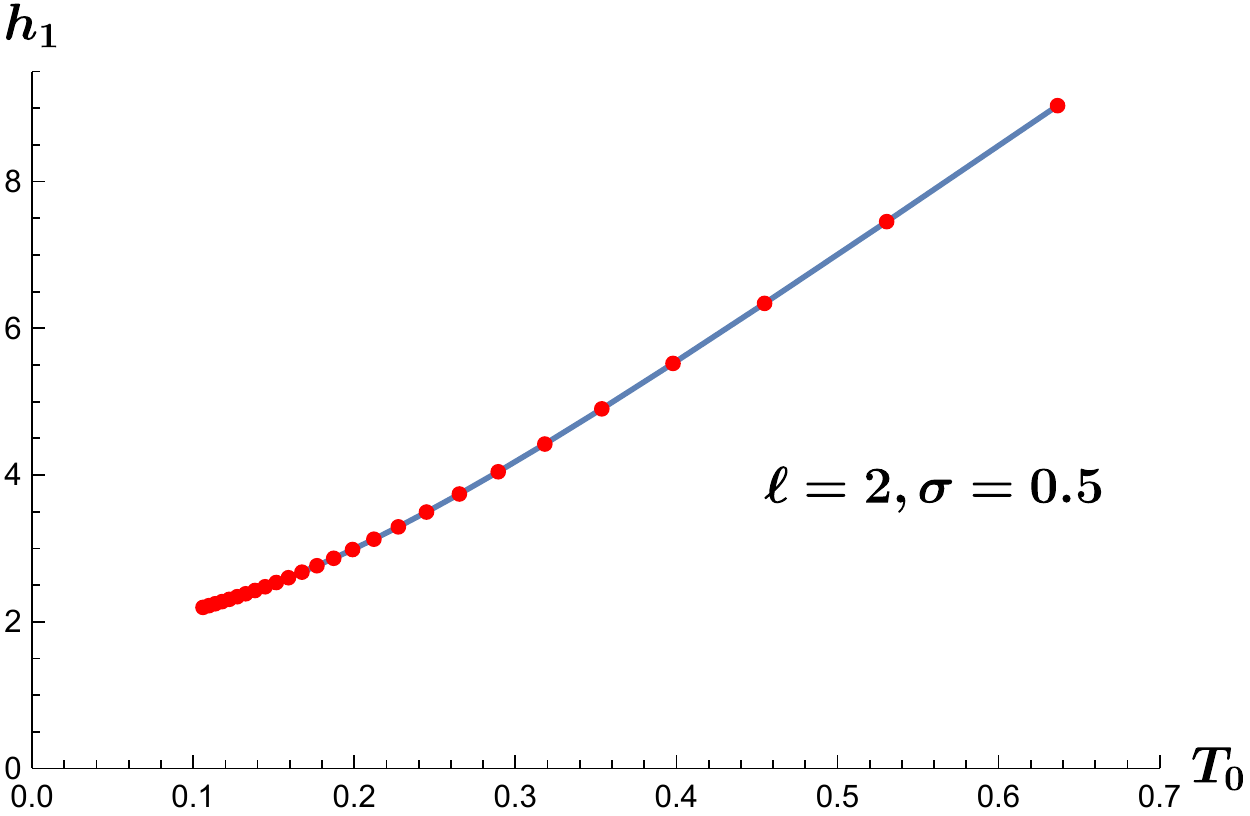}
  \caption{}
  \label{fig:h1pl}
\end{subfigure}%
\begin{subfigure}{.5\textwidth}
  \centering
  \includegraphics[width=0.9\linewidth]{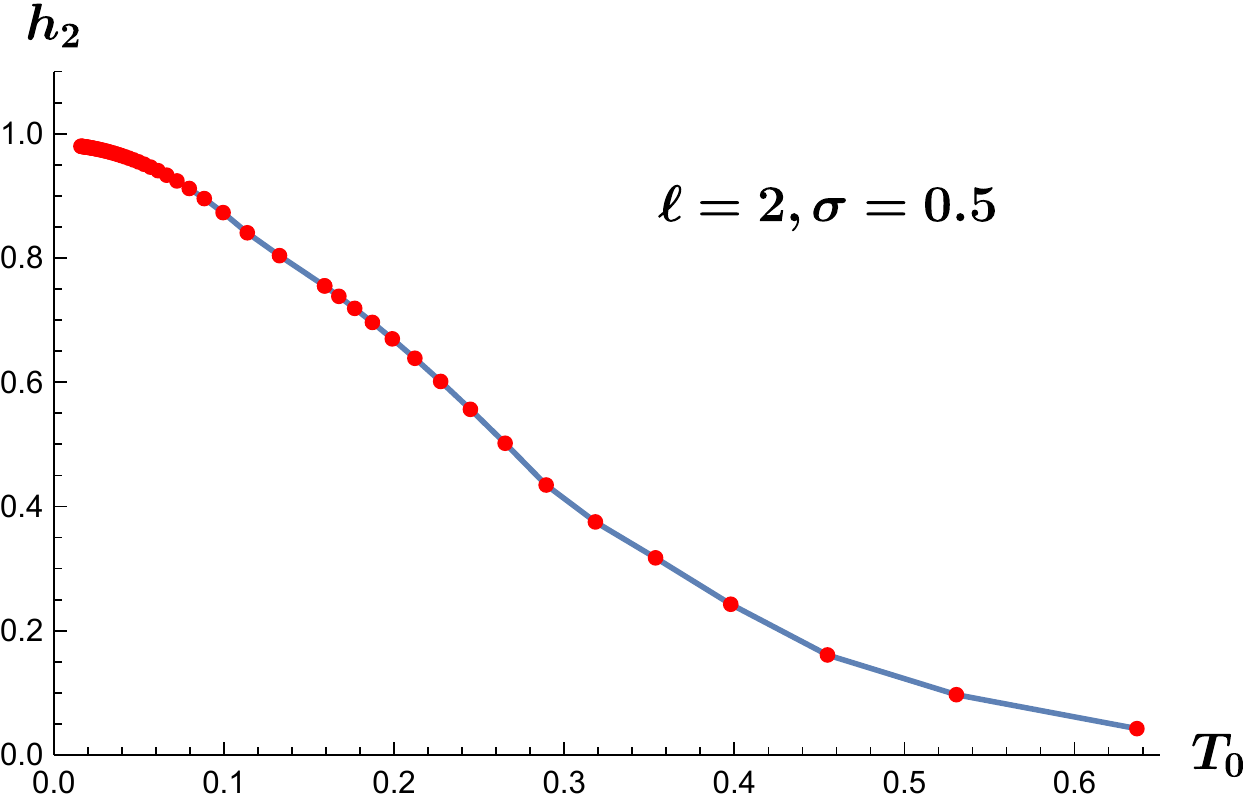}
\caption{}
  \label{fig:h2pl}
\end{subfigure}
\caption{The variation of the height $h_1$ and dip $h_2$ of $\Delta S_E$ (in fig.\ref{fig:delSpvst}) as a function of temperature.}
\label{fig:h1h2pl}
\end{figure}

The extent of the dip ($h_2$), and 
the maximum value of $\Delta S_E$ during the process ($h_1$), both are determined by 
the background temperature $T_0$, for a fixed width of the pulse. In fig.\ref{fig:h1h2pl}, we plot the variation of $h_1$ and $h_2$ with respect 
to background temperature. We note that $h_2$ decreases with temperature, while $h_1$ increases. It appears from fig.\ref{fig:h2pl}
that $h_2$ would perhaps tend to a finite value 
as $T_0 \rightarrow 0$. This suggests that 
the physical origin of this dip may be due to some non-thermal effect, which comes into play when 
propagating fluctuations of energy-momentum tensor tries to enter or exit the subsystem.
%
%

%
\subsection{Dissipative sound mode in higher dimensions}\label{ssec:sound4D}

We now move on to higher dimension, where the sound mode is dissipative. Unlike, $d=2$, in higher dimensions 
we have a derivative expansion, whose validity restricts our analysis only to long wavelengths i.e. $k/T \ll 1$. 
The HEE for these dissipative states are studied for a subsystem of length $\ell$, so that $T \ell \sim \mathcal O(1)$. 
Here, we will follow the general procedure outlined in section \ref{ssec:genpro}, which will closely resemble the analysis in section \ref{sound2D}. 

We will first perform our analysis for $d=3$, where the time-dependence of HEE is predictably decaying sinusoidal, following 
the form of the sound mode, but we find an interesting phase difference between them, which carries non-trivial information regarding the entanglement structure of the state. 
We then move on to $d=4$, where apart from this phase, we observe a 
subleading UV divergence in HEE for the sound mode. Such an additional UV divergence, was absent in $d=2,3$. Since, our analysis is partially numerical, we therefore, restrict our analysis only up to $d=4$, but the qualitative features of the phase shift and the additional UV divergence is expected to be straightforwardly extended to higher dimensions. 

\subsubsection{$d=3$}\label{sssec:2p1sm}
%
Our starting point of this analysis is the holographic dual 
to $2+1$D boundary relativistic fluid, which is given by the $3+1$D bulk metric \cite{VanRaamsdonk:2008fp,Haack:2008cp,Bhattacharyya:2008mz}
\begin{equation}\label{gravmet3d}
\begin{split}
ds^2&=\frac{2}{z^2}u_\mu dx^\mu dz-\frac{1}{z^2}f^{(3)}(z) u_\mu u_\nu dx^\mu dx^\nu+\frac{1}{z^2}P_{\mu\nu}dx^\mu dx^\nu\\
&+ \mathfrak{a} \frac{2\, \zh}{z^2}\, F^{(3)}(z) \sigma_{\mu\nu} dx^\mu dx^\nu + \mathfrak{b} \frac{1}{z} u_\mu u_\nu (\partial_\lambda u^\lambda) dx^\mu dx^\nu-\frac{1}{z} u^\lambda\partial_\lambda(u_\nu u_\mu)dx^\mu dx^\nu+{\cal O}(\partial)^2
\end{split}
\end{equation}
where, the projector $P_{\mu \nu}$ has the usual definition, and 
\begin{equation*}
\begin{split}
& \sigma^{\mu\nu}=\frac{1}{2}P^{\mu\alpha} P^{\nu\beta}(\partial_\alpha u_\beta+\partial_\beta u_\alpha)-\frac{1}{2}(\partial_\alpha u^\alpha)P^{\mu\nu}\\
& f^{(3)}(z)=1-\frac{z^3}{\zh^3}, \quad 
F^{(3)}(z) =\frac{1}{6}\left[\sqrt{3}\,\pi-2\sqrt{3}\, \tan^{-1}\frac{2\,\zh+z}{\sqrt{3}\,z}+3\log\frac{\zh^2+\zh\,z+z^2}{\zh^2}\right]
\end{split}
\end{equation*}
Also, compared to the original metric reported in \cite{VanRaamsdonk:2008fp,Haack:2008cp,Bhattacharyya:2008mz}, 
we have introduced two additional parameters which must be set to $\mathfrak{a} =1 , \mathfrak{b}=1$ so as to ensure that 
\eqref{gravmet3d} is a solution to Einstein's equations. 
The terms, whose coefficients are $\mathfrak{a}$ and $\mathfrak{b}$, will play an important role in our discussion, 
which is why we have chosen to track them through the calculations with the help of these two parameters.
We shall report all our results setting $\mathfrak{a} = 1 , \mathfrak{b}=1$, and refer to them only when necessary 
for contextual clarity. 

As usual, we denote the boundary coordinates with $x^\mu=\{t,x,y\}$, while the temperature $T$ \eqref{temp} and 
velocity $u_\mu= \gamma\{-1,v,0\}$ constitutes the fluid variables. The metric \eqref{gravmet3d} 
is a solution of Einstein's equation provided the fluid variables satisfy 
$
\partial_\mu T^{\mu\nu}=0
$
where, in this case, the constitutive relation reads 
\begin{equation}
T_{\mu\nu}= \mathcal P \left(\eta_{\mu\nu}+3\, u_\mu u_\nu\right)-2 \, \eta \, \sigma_{\mu\nu}+{\cal O}(\partial)^2.
\end{equation}
where the pressure density $\mathcal P = (4 \pi T / 3)^3$ and shear viscosity $\eta = (4 \pi T / 3)^2$.

\subsubsection*{Linearized sound-wave solutions}
Here, the fluid equations admit the following linearized sound wave solution
\begin{equation}\label{exp3dsound}
\begin{split}
\zh&=\zh_0 \left( 1 + \epsilon\  \exp[-i(\omega\, t-k\, x)]+ ~\mathcal O(\epsilon^2) \right) \\
v&=\epsilon\, v_1 \exp[-i(\omega\, t-k\, x)]+~ \mathcal O (\epsilon^2)
\end{split}
\end{equation}
with,
$$\omega=\pm\frac{1}{\sqrt{2}}k-\frac{i \zh_0}{6}k^2+{\cal O}(k^3), ~~v_1= \sqrt{2}+i\frac{\zh_0 k}{3} + \mathcal O(k^2)$$
Note that due to the imaginary piece in 
the dispersion relation, the fluctuations 
are damped exponentially by a factor of 
$\exp (- \lambda t)$, 
where the damping constant $\lambda = \frac{1}{6} \eta k^2 \zh_0^3$ is determined by shear viscosity\footnote {In the holographic calculation, $\eta$ is always determined by the temperature, and therefore, its presence in various computations becomes obscure. Here, we know that $\eta$ is the only source of dissipation in this problem.}.

\subsubsection*{Extremal surface in the background}
To implement the general procedure of calculating $\Delta S$ (see section \ref{ssec:genpro}), we have to calculate the extremal surface in the static background metric
\begin{equation}\label{eq:leadmet}
ds^2=\frac{1}{z^2}\left[-2\, dt\, dz-\left(1-\frac{z^3}{\zh_0^3}\right) dt^2+dx^2+dy^2\right]\,,
\end{equation}
This metric is in Eddington-Finklestein. Therefore, following \eqref{embedEFnob}, we can choose the ansatz for the extremal surface to be 
\begin{equation}\label{embed3Dz}
 \Gamma: ~x= x(z),~ t= t_b - h(z) ,~ z = z, ~ y = \tilde y, 
\end{equation}
where, 
\begin{equation}
 h(z) = \int_0^{z} \frac{dz'}{\left(1-\frac{z'^3}{\zh_0^3}\right)} =\frac{\zh_0}{6}\left(2\sqrt{3}\, \tan^{-1}\frac{2\,z+\zh_0}{\sqrt{3}\,\zh_0}+\log \frac{z^2+z\,\zh_0+\zh_0^2}{(\zh_0-z)^2}-\frac{\pi}{\sqrt{3}}\right)
\end{equation}
The ansatz leads to the following expression for $A_\Gamma^{(0)}$ (see \eqref{HEEeq})
\begin{equation}
 A_\Gamma^{(0)} = \int \sqrt{\mathcal{G}_0}\, dz\, dy=2\,L\int_0^{z_\star}\frac{dz}{z^2}\left[ \frac{1}{1-\frac{z^3}{\zh_0^3}}+x^\prime(z)^2\right]^\frac{1}{2}
\end{equation}
which upon extremization yields 
\begin{equation}\label{3Dextsur}
x^\prime(z)=\frac{\frac{z^2}{{z_\star}^2}}{\sqrt{1-\frac{z^3}{\zh_0^3}}\sqrt{1-\frac{z^4}{{z_\star}^4}}}
\end{equation}
Again, $z_\star$ is the turning point, which is determined in terms of the boundary subsystem length $\ell$. The integration 
in \eqref{3Dextsur} is performed numerically. 

\subsubsection*{Computation of $\Delta S_E$ for sound mode}
%

The change in the area of the extremal surface $\Gamma$ in \eqref{embed3Dz} due to linearized sound wave fluctuations, 
which captures the corresponding change in entanglement entropy, is given by (see \eqref{genformu})
\begin{equation}\label{3DswdelS}
\Delta S_E =\frac{A_\Gamma^{(1)}}{L}  =  \int_0^{z_\star} dz\, \sqrt{{\mathcal{G}_0}}\, {\mathcal G}_0^{\alpha\beta}{\mathcal G}^1_{\alpha\beta} =  \int_0^{z_\star} dz\, \mathfrak I(z)
\end{equation}
where, $\mathfrak I(z)$ has the following form
\begin{equation}\label{Iform}
\mathfrak I (z)= e^{-\lambda \, t(z)} \exp \left( {i\,k\left(\frac{t(z)}{\sqrt{2}}+x(z)\right)} \right) 
\left( \mathfrak I_R(z) +i\, \mathfrak I_I(z) \right)
\end{equation}
where,
\begin{equation}\label{GFexpand3D}
\begin{split}
\mathfrak I_I&=k\left(-(\mathfrak{a}-\mathfrak{b})\frac{\sqrt{2}}{z}+\frac{2\,\zeta_0}{3\,{z_\star}^2}+{\cal O}(z)\right)+{\cal O}(k^2)\\
\mathfrak I_R&=\left(\frac{2\sqrt{2}}{{z_\star}^2}+{\cal O}(z)\right)+{\cal O}(k^2)
\end{split}
\end{equation}
We have expanded $\mathfrak I_I$ and $\mathfrak I_R$ in $z$, about the asymptotic boundary, to demonstrate the following interesting feature of the integrand $\mathfrak I(z)$. 

Let us recall that since $\Delta S_E$ captures the change in 
HEE compared to the thermal background value, we do not expect any leading order divergence in $\Delta S_E$, but there may 
be subleading divergences. Also, since we are in d=3, a subleading divergence would mean a logarithmic divergence. 
Clearly, in \eqref{GFexpand3D} we see that there is a term in $\mathfrak I_I$ which is $\mathcal O \left( 1/z \right)$. This term
may lead to a potential UV logarithmic divergence upon performing the integration in \eqref{3DswdelS}. However, 
the term is proportional to $(\mathfrak{a}-\mathfrak{b})$ and we know for the fluid metric to satisfy the Einstein's 
equation we must have $\mathfrak{a}=1$ and $\mathfrak{b}=1$. Thus consistency with the right fluid-gravity metric necessitates 
the vanishing of this potentially divergent term. We should note that this cancellation occurs only at the linear order in $k$. 
This is the order up to which our calculation is reliable since we have considered only first order in derivative expansion.
It is possible that at higher order this subleading logarithmic divergence exists with some universal coefficient. 
As we shall see below, this magical cancellation does not happen in $d=4$, where a subleading UV divergence is observed. 
\begin{figure}
\centering
\begin{subfigure}{.5\textwidth}
  \centering
  \includegraphics[width=0.9\linewidth]{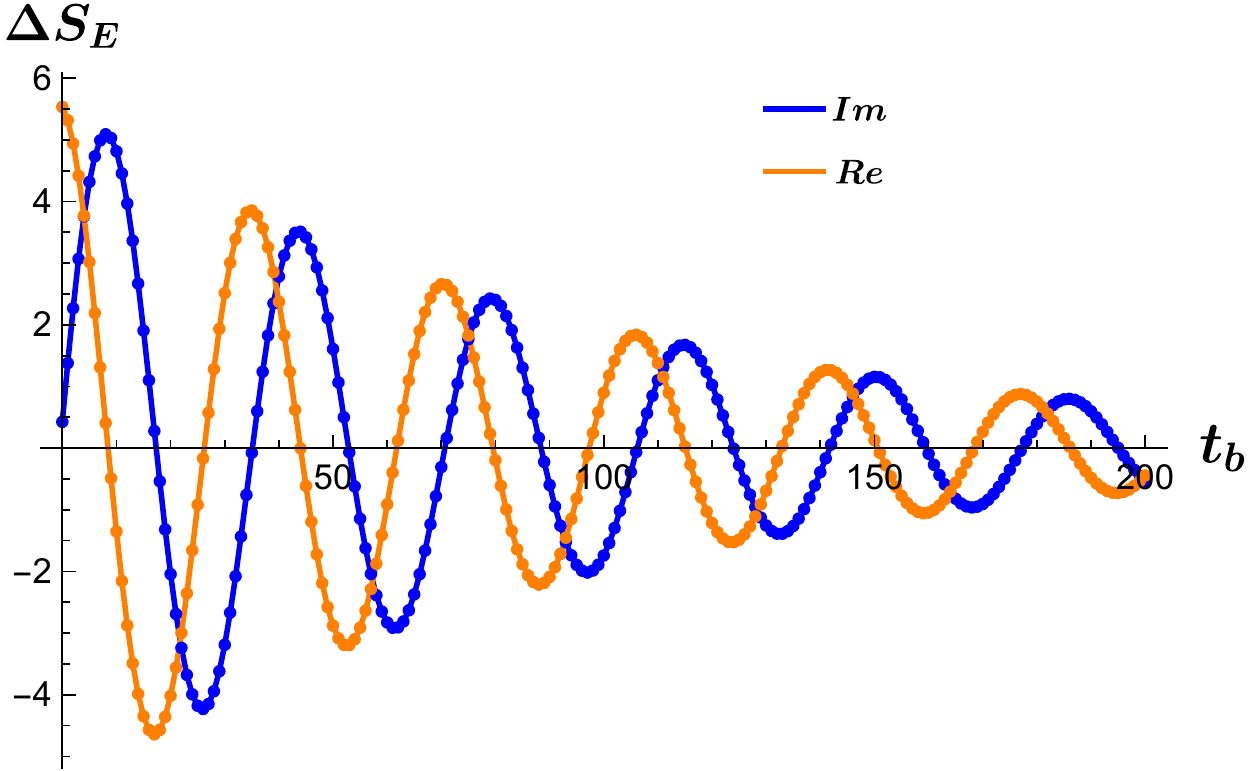}
  \caption{}
  \label{fig:Desvat3Dsw}
\end{subfigure}%
\begin{subfigure}{.5\textwidth}
  \centering
  \includegraphics[width=0.9\linewidth]{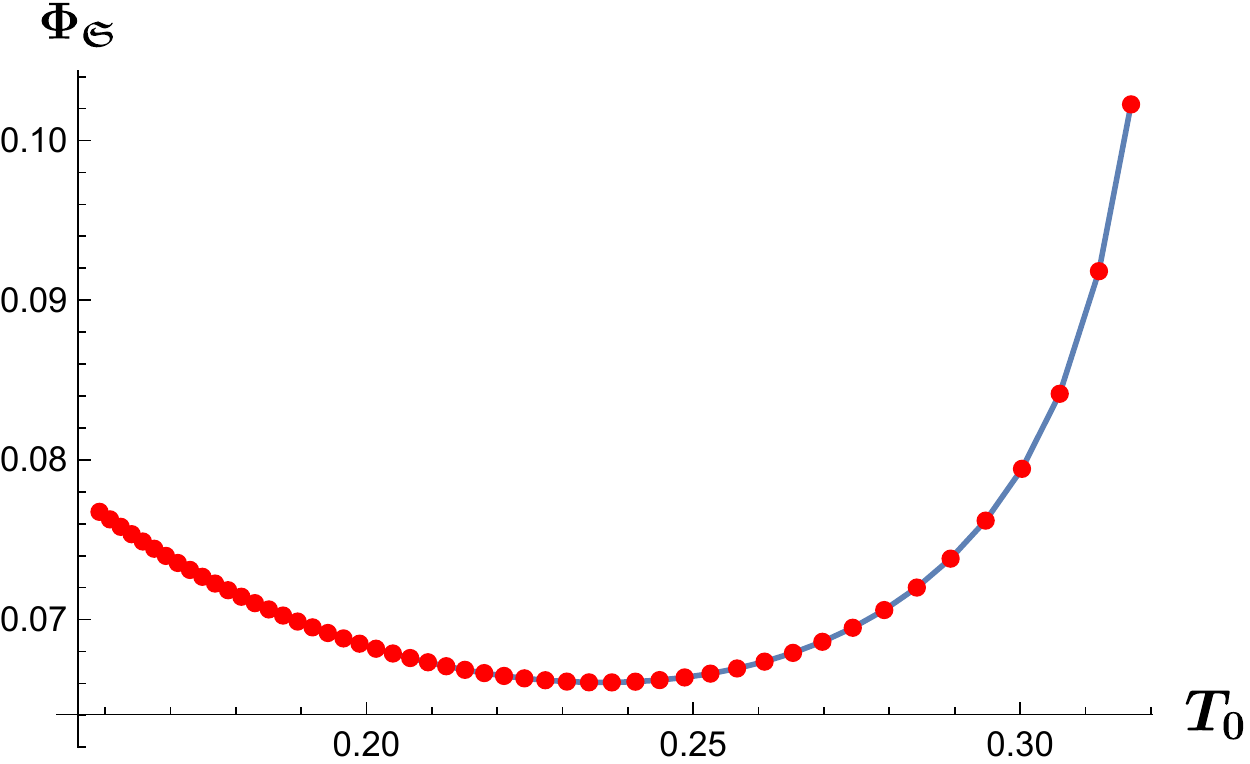}
\caption{}
  \label{fig:phase3D}
\end{subfigure}
\caption{In (a) we plot the variation of $\Delta S_E$ with time $t_b$ for the sound 
mode in $d=3$. This variation is trivially damped harmonic 
as predicted by \eqref{delSform3d}. In (b) we plot the phase of the amplitude 
$\Phi_{\mathfrak S}$ as a function of background temperature $T_0$. 
}
\label{fig:3dsoundplot}
\end{figure}

From the structure of \eqref{Iform} and the embedding equation \eqref{embed3Dz}, it is immediately 
clear that $\Delta S_E$ has a straightforward dependence on $t_b$, which factors out of the integral. 
In fact upon performing the integration 
in \eqref{3DswdelS}, we should have the following schematic form
\begin{equation}\label{delSform3d}
\begin{split}
 \Delta S_E (t_b) &=   e^{-\lambda \, t_b} \exp \left( {i\,k\frac{t_b}{\sqrt{2}}} \right) \mathfrak S (k,\ell,T), \\
 \text{where}, ~~ \mathfrak S (k,\ell,T) &= \mathfrak S^{(0)} (\ell,T) + k ~\mathfrak S^{(1)} (\ell,T) + \mathcal O (k^2). 
 \end{split}
\end{equation}
The real and imaginary parts of  $\Delta S_E (t_b)$ has been plotted in fig.\ref{fig:Desvat3Dsw}, which is clearly 
damped harmonic due to form \eqref{delSform3d}. The phase of $\mathfrak S$, which we denote by $\Phi_{\mathfrak S}$, represent a phase shift in $\Delta S_E$ compared to the original sound wave perturbations \eqref{exp3dsound}. 
The variation of this phase with respect the $T\ell$, for a fixed $k$, has been shown in fig.\ref{fig:phase3D}. We find that
this phase has a weak dependence on temperature, 
and exhibits a growing trend towards high temperature. 

\subsubsection{$d=4$}\label{sssec:3p1sm}

The calculation in $d=4$ proceeds in parallel with that in $d=3$. 
We consider the following fluid-gravity metric \cite{Bhattacharyya:2007vjd}, corrected up to first order in derivative expansion  in $4+1$ bulk dimension
\begin{equation}\label{gravmet4d}
\begin{split}
ds^2&=\frac{2}{z^2}u_\mu dx^\mu dz-\frac{1}{z^2}f(z) u_\mu u_\nu dx^\mu dx^\nu+\frac{1}{z^2}P_{\mu\nu}dx^\mu dx^\nu\\
&+\mathfrak{a}\frac{2\, \zeta}{z^2}\, F(z) \sigma_{\mu\nu} dx^\mu dx^\nu+\mathfrak{b}\frac{2}{3\,z} u_\mu u_\nu (\partial_\lambda u^\lambda) dx^\mu dx^\nu-\frac{1}{z} u^\lambda\partial_\lambda(u_\nu u_\mu)dx^\mu dx^\nu+{\cal O}(\partial)^2
\end{split}
\end{equation}
where,
\begin{equation}
\begin{split}
f(z)=1-\frac{z^4}{\zeta^4}, ~\text{and}~  
F(z)=\frac{1}{4}\left(\log\frac{(z+\zeta)^2(z^2+\zeta^2)}{\zeta^4}-2\tan^{-1}\frac{\zeta}{z}+\pi\right)
\end{split}
\end{equation}
Just like in $d=3$, we have introduced two constant parameters $\mathfrak{a}$ and $\mathfrak{b}$ in the metric, which must 
take the values $\mathfrak{a}=1$ and $\mathfrak{b}=1$, so as to ensure \eqref{gravmet4d} solves Einstein's equation. 

We denote the boundary coordinates with $x^\mu=\{t,x,y_1,y_2\}$ and express the two fluid variables as
\begin{equation}
u_\mu= \gamma\{-1,v,0,0\},\quad T=\frac{1}{\pi\,\zeta}.
\end{equation}
The fluid variables in \eqref{gravmet4d} must satisfy the constraint equation 
\begin{equation}\label{eq:Tmunu}
\partial_\mu T^{\mu\nu}=0.
\end{equation}
where,
\begin{equation}
T_{\mu\nu}=\frac{1}{\zeta^4}\left(\eta_{\mu\nu}+4\, u_\mu u_\nu\right)-\frac{2}{\zeta^3}\sigma_{\mu\nu}+{\cal O}(\partial)^2
\end{equation}
We consider the linearized sound wave solution 
\begin{equation}\label{exp4dsound}
\begin{split}
\zh&=\zh_0 \left( 1 + \epsilon\  \exp[-i(\omega\, t-k\, x)]+ ~\mathcal O(\epsilon^2) \right) \\
v&= \epsilon\, v_1 \exp[-i(\omega\, t-k\, x)]+~ \mathcal O (\epsilon^2) \\
\text{with},~~ \omega &=\pm\frac{1}{\sqrt{3}}k-\frac{i \zeta_0}{6}~k^2 + {\mathcal O}(k)^3,
~v_1 = \sqrt{3} + i\frac{k \zeta_0}{2} + {\mathcal O}(k)^2.
\end{split}
\end{equation}
Note that, in $d=4$ the damping constant 
$\lambda = \frac{1}{6} \eta k^2 \zh_0^4$. 
As in $d=3$, the required surface is to be extremized in the background geometry 
\begin{equation}\label{4dbkmet}
ds^2=\frac{1}{z^2}\left(-2\, dt\, dz-\left(1-\frac{z^4}{\zeta_0^4}\right) dt^2+dx^2+dy_1^2+dy_2^2\right)\,.
\end{equation}
The ansatz for the extremal surface  in Eddington-Finkelstein coordinates is again 
\begin{equation}\label{embed3Dz2}
 \Gamma: ~x= x(z),~ t= t_b - h(z) ,~ z = z, ~ y_i = \tilde y_i, 
\end{equation}
where
\begin{equation}
h(z)= \int_0^z \frac{dz'}{\left(1-\frac{z'^4}{\zeta_0^4}\right)} = \frac{\zeta_0}{2} \tan^{-1}\frac{z}{\zeta_0}-\frac{\zeta_0}{4}\log[\zeta_0-z]+\frac{\zeta_0}{4}\log[\zeta_0+z].
\end{equation}
and 
\begin{equation}
x^\prime(z)=\frac{\frac{z^3}{{z_\star}^3}}{\sqrt{1-\frac{z^4}{\zeta_0^4}}\sqrt{1-\frac{z^6}{{z_\star}^6}}},
\end{equation}
which is numerically integrated to obtain the necessary minimal surface. 
With the help of this extremal surface, 
the change in HEE due to the sound mode \eqref{exp4dsound} is evaluated  
\begin{equation}
\Delta S_E = \frac{ A_\Gamma}{L^2} =  \int^{z_\star}_0 dz~ \mathfrak I (z),
\end{equation}
where, $\mathfrak I(z)$ has the following schematic form
\begin{equation}
\mathfrak I (z)= e^{-\lambda t(z)} \exp \left[ {i\,k\left(\frac{t(z)}{\sqrt{3}}+x(z)\right)} \right] 
\left(\mathfrak I_R(z)+i\, \mathfrak I_I(z)\right)
\end{equation}
where, near the asymptotic boundary ($z\rightarrow 0$), we have 
\begin{equation}
\begin{split}
\mathfrak I_I&=k\left(-(2\mathfrak{a}-\mathfrak{b})\frac{2}{\sqrt{3}z^2}+\frac{\zeta_0}{{z_\star}^3}+{\mathcal O}(z)\right)+{\cal O}(k^2)\\
\mathfrak I_R&=\left(\frac{2\sqrt{3}}{{z_\star}^3}+{\cal O}(z)\right)+{\cal O}(k^2)
\end{split}
\end{equation}
Recalling that $\mathfrak{a}=1, \mathfrak{b}=1$, we conclude that there exists a $\mathcal O(1/z^2)$ term 
in the integrand $\mathfrak I(z)$. Once integrated, this term will give rise to a term in $\Delta S_E$ which is 
proportional to $1/\euv$, where $\euv \rightarrow 0$ is the UV cut-off. Note that this divergence
is subdominant compared to the `area law' divergence \cite{Nishioka:2009un}, which for $d=4$ is proportional to 
$1/\euv^2$ \footnote{The presence of this UV divergence in $\Delta S_E$ definitely puts a question 
mark on our perturbative approach. But, since this divergence is subdominant compared to the leading `area law', 
our perturbation theory is perhaps not completely invalidated.}. 
We should emphasize that our observation is limited to $\mathcal O(k)$, and it is possible that 
there are other subleading divergences at higher order in the derivative expansion. 

The integral in $\Delta S_E$ can be performed numerically only after extracting out the divergence. The integrated answer has the final schematic form 
\begin{equation}\label{delSform4d}
\begin{split}
 \Delta S_E (t_b) &=   e^{-\lambda \, t_b} \exp \left( {i\,k\frac{t_b}{\sqrt{3}}} \right) 
 \left( - \frac{2 k}{\sqrt{3} \euv} + \mathfrak S (k,\ell,T)\right), \\
 \text{where}, ~~ \mathfrak S (k,\ell,T) &= \mathfrak S^{(0)} (\ell,T) + k ~\mathfrak S^{(1)} (\ell,T) + \mathcal O (k^2). 
 \end{split}
\end{equation}
Here, the function $\mathfrak S (k,\ell,T)$ is UV finite and has the same qualitative form as in $d=3$ (see fig.\ref{fig:3dsoundplot}). Also, recall that we have rescaled HEE by a factor of $L^2$. Hence, this UV divergent 
piece in the full HEE is proportional to $(k L^2 / \euv) e^{-\lambda \, t_b} \exp \left( {i\,k\frac{t_b}{\sqrt{3}}} \right)$, 
which is subdominant compared to the area law divergence proportional to $L^2 / \euv^2$. Clearly, the subleading divergence decays with time and disappears in the final equilibrium state.

\section{Discussions}\label{sec:disco}

In this paper, we have studied holographic entanglement entropy for 
two varieties of fluid configurations. At first, we have considered a
stationary configuration where the fluid moves with a constant 
velocity. Here we find that HEE generically increases with an increase in 
fluid velocity when the other parameters are held fixed. Also, as the fluid approaches its relativistic 
upper bound $v \rightarrow 1$, the regulated HEE diverges. From our analysis, it appears 
that this divergence is proportional to a factor of $\gamma$ in $d=2,3$. 
This is perhaps a universal behaviour for strip-shaped subsystems which we consider here. 
Admittedly, we do not have a clear understanding of this additional divergence.  
However, we would like to point out that a similar divergence was also present in the earlier perturbative results of \cite{Blanco:2013joa}, although the structure of the divergence is slightly different in our non-perturbative treatment. 

In this context, we also study holographic mutual information between two non-overlapping subsystems of the same size, 
separated by a distance $x$. When the fluid velocity is zero ($v=0$), it is known from previous work \cite{Fischler:2012uv} that this mutual information vanishes at large $x$, while it is non-zero at small $x$. There is a `phase-transition'-like effect for mutual information at some critical distance $x_c$. 
For our steady state fluid flows, we find that even when the subsystems are close by ($x < x_c$), there exists 
a critical value of fluid velocity $v_c$ above which mutual information vanishes 
and remains zero for higher velocities. This occurs due to the flip of 
dominance between extremal surfaces, similar to what is observed 
when at $v=0$, the separation between the subsystems is increased. 
This critical velocity exists for all values of $x < x_c$. 
We find that $v_c$ increases with a decrease in the separation
between the two subsystems, and $v_c \rightarrow 1$ as the separation approaches zero. 
This result follows from an analytical formula in $d=2$ (see \eqref{MI2dexct}), while it has been verified 
numerically in $d=3$ (see fig.\ref{fig:MIhighD}). Again, we think, the qualitative aspects of this observation 
straightforwardly generalize to higher dimensions. 

Our computation of entanglement measures for the steady-state fluid flow has a few immediate generalizations. 
In our work, we have always assumed that the fluid velocity is along the short edge of the strip. This may be generalized to a general angle, giving rise to a richer structure of HEE. It is tempting to speculate that, as we vary this angle, there may be a phase-transition-like effect as observed in \cite{Narayan:2012ks}. Also, along this direction, generalizing our work to other shapes for the subsystem should also be interesting. 

We then proceed to study HEE for the sound mode excitation of the fluid. Firstly, in $d=2$ the sound mode is non-dissipative. The variation 
of HEE with respect to boundary time is, in a way trivial since it has a predictable sinusoidal behaviour mimicking that 
of the fluid variables in the linearized approximation. However, it is observed that the amplitude 
of HEE has a very interesting dependence on the momentum $k$ of the sound waves. 
This behaviour is quasi-periodic, with HEE vanishing at specific values of $k$, which are determined 
by the background temperature (see fig.\ref{fig:delsvsksw}). 
The sound modes may be superimposed to form a traveling Gaussian wave-packet, which is a positive pressure 
or energy density pulse. While this pulse moves across the subsystem of interest, 
HEE generically experiences a jump. However, we find that 
when the pulse enters or exits the subsystem, there is a curious dip in HEE (see fig.\ref{fig:delSpvst}). 
The pulse may also be thought of as a `semi-local' quench with the energy-momentum tensor,
and similar effects on HEE may also be visible for a non-thermal background as well. 
It would be very interesting to investigate the physical 
origin of this dip in future work. 

Finally, we have considered the sound mode in $d=3,4$, where it is dissipative\footnote{There is a significant recent interest in the entanglement structure of non-unitary systems (see \cite{Goto:2022fec} for a recent work). This part of our work is definitely in line with such developments.}. The fluid fluctuations are damped due to the shear viscosity, and at late times 
these configurations settle down to equilibrium. Due to our linearized 
approximation, the time dependence of HEE follows this damped oscillatory 
behaviour of the fluid variables (see fig.\ref{fig:3dsoundplot}). But, the amplitude of these oscillations 
is significantly interesting due to a phase which is physically relevant.  
The most noteworthy feature in this calculation is the presence of 
an additional UV divergence in $d=4$, which is subleading compared to the `area law divergence'. 
This divergence is dynamical in the sense that its coefficient is exponentially damped at late times and therefore 
disappears in the final equilibrium state. In $d=3$, potentially, 
there was the possibility of such a subleading logarithmic divergence, but it is 
absent due to a miraculous cancellation at first order in derivative expansion. The possibility of 
the existence of this divergence in $d=3$ when higher order derivatives are taken into account remains open. 
The physical origin of this divergence in $d=4$ is not very clear. 
Dissipation is clearly not the reason, as it is absent for the $d=3$ sound mode\footnote{The divergence also appears to be absent for diffusive effects in $d=4$ which can be seen from the asymptotic fall off of the metric functions in \cite{Banerjee:2008th,Erdmenger:2008rm}.}. It is possible that this is some artifact of the derivative expansion. 
In future investigations, it would definitely be very interesting to develop a better understanding of this mysterious divergence. 

In our analysis of the sound mode, we have considered the fluctuations as a linearized excitation 
over a static background. It would be very interesting to turn on 
a background fluid velocity for this question. The interplay between the 
directions of the background fluid velocity and that of the sound mode would definitely 
have important consequences for holographic entanglement. 

The fluid states are generally very complex macroscopic states whose underlying entanglement structure would be extremely difficult 
to estimate using quantum field theoretic techniques. For conformal fluids, the holographic set up provides us with a unique opportunity to compute such quantities relatively easily. Our HEE results here, are primarily of theoretical interest, and we believe, it adds to our understanding of the interplay between quantum information measures and macroscopic dynamics. With this in view, extending our line of investigation to holographic superfluids would be particularly interesting. Since our calculations rely heavily on holographic methods, this is perhaps its main limitation. Technically our results are only valid for CFTs admitting a holographic dual. However, we would expect some of the qualitative features of our result to be valid beyond the purview of holography. 
 Also, the fluid-gravity solutions we have used are approximate solutions written in a derivative expansion. Hence our results of HEE for the dissipative sound mode in $ d = 3, 4$ is limited by this approximation. In all our investigations we have used a strip-like subsystems, and generalization of this assumption may also lead to more interesting results in this context. Extending our entanglement entropy investigation to more complicated fluid flows beyond the sound mode (such as turbulent flows) would also be very interesting future work. 

\acknowledgments 
%
We would like to thank Anirudh Deb for initial collaboration and many useful discussions. 
We are also particularly grateful to Anirban Dinda and Nilakash Sorokhaibam for many insightful discussions. 
We would also like to thank Sayantani Bhattacharyya, Diptarka Das, Bobby Ezhuthachan, Sayan Kar, Apratim Kaviraj, S. Pratik Khastgir, Arnab Kundu, Nilay Kundu, R. Loganayagam,  Sabyasachi Maulik, Kannabiran Seshasayanan, Vishwanath Shukla for several useful discussions. 
We are also extremely thankful to Shankhadeep Chakrabortty, Anirudh Deb and Tadashi Takayanagi for their valuable comments on the draft of our paper. JB and SKD would like to acknowledge hospitality at NISER Bhubaneshwar, during the workshop titled `Regional strings meeting 2022', where some of the results reported in this paper were first presented. PB would like to acknowledge hospitality at IIT Kharagpur during the course of this work. JB would like to acknowledge support from the Institute Scheme for Innovative Research and Development (ISIRD), IIT Kharagpur, Grant No. IIT/SRIC/PH/RFL/2021-2022/091. PB would like to acknowledge the support provided by the grant CRG/2021/004539.
%
%
\appendix

\section{Holographic entanglement in Eddington Finklestein coordinates}\label{app:EFcoord}
%
Here we briefly discuss the procedure to obtain HEE
in Eddington-Finklestein (EF) coordinates. 
Since the horizon of the black brane is regular in these coordinates, their use 
is essential for performing a well-defined derivative expansion 
in the fluid gravity correspondence. Hence, the extremal surface in 
EF coordinates must be used in the computations of section  \ref{sec:hydromodes}. 

The boosted black brane metric in $d+1$ dimensional EF coordinates takes the form 
\begin{equation}\label{EFbb}
 ds^2_{\text{EF}} = \frac{1}{z^2} \left( - f(z) u_\mu u_\nu dx^\mu dx^\nu + 2 u_\mu dx^\mu dz + P_{\mu \nu} dx^\mu dx^\nu \right)
\end{equation}
where $u \cdot u = - 1$, and $P_{\mu \nu} = \eta_{\mu \nu} + u_{\mu} u_{\nu}$ is the projector orthogonal to $u_{\mu}$. The function $f(z)$ is identical to that appearing in \eqref{schboost}. In \eqref{schboost}, we have considered the boost to be in the $x$-direction, correspondingly we should have 
$u_{\mu} = \gamma \{-1,v,0,0, \cdots\}$, with $\gamma = (1-v^2)^{-1/2}$.   

The EF coordinates in \eqref{EFbb} are related to the Schwarzschild coordinates 
in \eqref{schboost}, by the following transformations 
\begin{equation}\label{EFtoSch}
 t = \tilde t ~+ ~\int_{0}^z \frac{\gamma}{f(z')} dz' , ~~ x = \tilde x ~ + ~\int_{0}^z \frac{v \gamma }{f(z')} dz' ,
\end{equation}
where $\tilde t$ and $\tilde x$ are respectively the time coordinate and space coordinate in EF coordinates, the $\tilde x$-coordinate being in the direction of the boost.  

Note that, both these coordinates reduce to the same Minkowski coordinates at the boundary of AdS. 
This ensures there is no geometrical transformation in the specification of the boundary subsystem for calculating HEE, as we move from the Schwarzschild
to the EF coordinates. However, the bulk extremal surfaces are different and are related to each other through the transformations \eqref{EFtoSch}, 
implemented on the embedding coordinates. 

Equipped with the transformation \eqref{EFtoSch}, we can convert the extremal surfaces for the boosted 
black brane in Schwarzschild coordinates \eqref{embedgend} (see fig.\ref{fig:extreme2D}), to those in EF coordinates \eqref{EFbb}.
This procedure is particularly useful for the zero boost case ($v=0$), when it is considerably easier to compute the extremal surface 
in Schwarzschild coordinates where it lies on a single bulk time-slice. 

For $v=0$, the extremal surface for the strip-subsystem in Schwarzschild coordinates takes the form 
\begin{equation}\label{embedSchnob}
 \Gamma_{S}: ~x= x(\phi),~ t = t_b,~ z = z_\star \cos \phi, ~ y_i = \tilde y_i. 
\end{equation}
where $t_b$ is the boundary time slice, on which the subsystem-$A$ is located. The non-trivial 
embedding function $x(\phi)$ is known exactly for the BTZ black hole, while in higher dimensions, 
it is given by an integral which may be straightforwardly computed numerically. For $v = 0$, 
the transformations \eqref{EFtoSch} reduced to 
\begin{equation}\label{EFtoSchnob}
 t = \tilde t ~+ ~\int_{0}^z \frac{1}{f(z')} dz' ,~~ x = \tilde x  ,
\end{equation}
Hence the embedding equations for extremal surface \eqref{embedSchnob} in EF coordinates is given by 
\begin{equation}\label{embedEFnob}
 \Gamma_{S}: ~\tilde x = \tilde x(\phi),~ \tilde t = t_b - \int_{0}^{z_\star \cos \phi} \frac{\gamma}{f(z')} dz' ,~ z = z_\star \cos \phi, ~ y_i = \tilde y_i. 
\end{equation}
The result \eqref{embedEFnob}, has been extensively used in section \ref{ssec:sound4D}, where the exact form of the zero boost extremal surface in EF coordinates was necessary to compute linear corrections to HEE corresponding to the sound mode.

\bibliographystyle{JHEP}
\bibliography{Entangled_hydro}
\end{document}